\documentclass[
twocolumn,
showkeys,
superscriptaddress,
longbibliography,
nofootinbib,
bibnotes,
nobibnotes,
amsmath,amssymb,
aps,
prd,
10pt]{revtex4-1}

\usepackage{hyperref}
\usepackage{tikz}
\usepackage{graphicx}
\usepackage{xcolor, soul}
\usepackage{dcolumn}
\usepackage{bm}
\usepackage{color}
\usepackage{enumitem}
\usepackage{mathrsfs}
\usepackage{cleveref}
\usepackage{orcidlink}
\usepackage{ulem}
\usepackage{multirow}
\usepackage[scr=boondox]{mathalfa}
\usepackage{subfig}
\usepackage{accents}

\begin{document}

    \title{Extreme mass ratio inspirals into black holes surrounded by matter:\\ Resonance crossings}

    \author{Michal Straten\'y\,\orcidlink{0009-0006-7623-3689}}
    \email{strateny.m@gmail.com}
    \affiliation{Astronomical Institute of the Czech Academy of Sciences, Bo\v{c}n\'{i} II 1401/1a, CZ-141 00 Prague, Czech Republic}
    \affiliation{Institute of Theoretical Physics, Faculty of Mathematics and Physics, Charles University, V Hole\v{s}ovi\v{c}k\'{a}ch 2, CZ-180 00 Prague, Czech Republic}

    \author{Georgios Lukes-Gerakopoulos\,\orcidlink{0000-0002-6333-3094}}
    \email{gglukes@gmail.com}
    \affiliation{Astronomical Institute of the Czech Academy of Sciences, Bo\v{c}n\'{i} II 1401/1a, CZ-141 00 Prague, Czech Republic}

    \author{Ond\v{r}ej Zelenka\,\orcidlink{0000-0003-3639-1587}}
    \email{ondrej.zelenka@asu.cas.cz}
    \affiliation{Astronomical Institute of the Czech Academy of Sciences, Bo\v{c}n\'{i} II 1401/1a, CZ-141 00 Prague, Czech Republic}
    
\begin{abstract}
        The forthcoming space-based gravitational-wave observatory Laser Interferometer Space Antenna (LISA) should enable the detection of Extreme Mass Ratio Inspirals (EMRIs), in which a stellar-mass compact object gradually inspirals into a supermassive black hole while emitting gravitational waves. Modeling the waveforms of such systems is a challenging task, requiring precise computation of energy and angular momentum fluxes as well as proper treatment of orbital resonances, during which two fundamental orbital frequencies become commensurate. In this work, we perform a systematic comparison of fluxes derived from three approaches: the quadrupole formula, post-Newtonian approximations, and time-domain solutions of the Teukolsky equation. We show that quadrupole-based fluxes remain in good agreement with Teukolsky results across a broad range of orbital configurations, including perturbed orbits. Building on these insights, we explore the dynamical impact of resonance crossings within the adiabatic approximation. By introducing novel numerical methods, we reduce computational costs and uncover diverse resonance-crossing behaviors. These results contribute to the effort to understand theoretically and model adequately resonance crossings during an EMRI.
\end{abstract}

\maketitle

\section{Introduction}\label{sec:Intro}
        The existence of gravitational waves (GWs), predicted by Einstein as a consequence of general relativity~\cite{Einstein1918}, were supported for decades by observing  electromagnetic waves emitted by inspiraling binary pulsar systems, like the famous PSR~B1913+16~\cite{Hulse1994}. This changed with the first direct detection, GW150914, by Laser Interferometer Gravitational-Wave Observatory (LIGO) on 14 September 2015~\cite{Abbott2016}, which inaugurated the era of GW astronomy~\cite{Bailes2021}. Since then, ground-based interferometers (LIGO, Virgo~\cite{Accadia2012}, and KAGRA~\cite{Akutsu2021}) have reported more than one hundred compact binary coalescences~\cite{Abbott2019,Abbott2021,Abbott2023}, with many more anticipated~\cite{Abbott2020}. These observatories are sensitive to the $Hz$ to $kHz$ frequency band and have been able to detect until now only mergers of comparable-mass binaries.

        To probe lower frequency bands space-based detectors are needed. The millihertz band, in particular, will enable observations of much longer events and more massive binary systems than the currently observed ones. For example, the Laser Interferometer Space Antenna mission, led by the European Space Agency~\cite{AmaroSeoane2017,Colpi2024}, is designed to survey a rich source population, including supermassive black hole (SMBH) mergers~\cite{Klein2016}, the long inspiraling phase of stellar mass binaries~\cite{Sesana2016}, and stochastic and cosmological GW backgrounds~\cite{Caprini2016,Tamanini2016,Bartolo2016}. Among the most compelling targets are extreme mass ratio inspirals (EMRIs)~\cite{Babak2017}, in which a stellar-mass compact object slowly inspirals into a SMBH, emitting richly structured GWs that encode the strong-field geometry.

        Namely, an EMRI consists of a compact object with mass $\mu$ of order $10^{0}$ up to $10^{2} M_\odot$ orbiting a galactic-center SMBH with mass $M$ of order $10^{5}$ up to $10^{8} M_\odot$~\cite{AmaroSeoane2017,Babak2017}. The extreme disparity in the mass ratio
        \begin{equation}\label{eq: mass ratio}
            q=\frac{\mu}{M}\ll 1, 
        \end{equation}
        leads to a slow inspiral that can remain in the band for months to years in the millihertz range~\cite{Berry2019,AmaroSeoane2017,Babak2017,Colpi2024}. The prolonged relativistic motion, often on inclined and eccentric orbits, imprints detailed signatures that can map the spacetime near the SMBH and probe strong-field gravity~\cite{Destounis2021,Destounis2020}, inform the dynamics of galactic nucleus~\cite{Barausse2014,Cardoso2022,Berry2019}, and yield precise measurements of black hole parameters~\cite{Berry2019}.

        Extracting this information requires accurate and computationally efficient waveform models for matched filtering~\cite{Helstrom1968}. Full numerical relativity simulations are infeasible for EMRIs~\cite{Lousto2023}. Self force methods provide a systematic framework~\cite{Barack2009,Pound2021}, but remain costly for large surveys. In this work we adopt an adiabatic approach in which the secondary follows a sequence of geodesics on a perturbed background, guided by precomputed energy and angular momentum fluxes. Our focus is on orbital resonances, where commensurabilities between fundamental frequencies modify dissipative dynamics and GW emission, potentially inducing waveform dephasing that can bias inference if left unmodeled~\cite{Speri2021,Berry2016}. Building on previous work~\cite{Strateny2023a}, we compare fluxes near resonances and analyze inspirals that traverse them, assessing the implications for waveform models in the LISA era.

        The rest of the article is organized as follows. In Section~\ref{sec: Section2}, we introduce the metric of the system under consideration together with the numerical methods employed for the computation of gravitational-wave fluxes. Section~\ref{sec: Section3} is devoted to a detailed comparison of these methods. In Section~\ref{sec: Section4}, we describe the adiabatic approximation adopted for modeling inspirals, and finally, in Section~\ref{sec: Section5}, we present results for adiabatic inspirals. Finally, Sec.~\ref{sec: Section6} summarizes the key findings of this work.

        We work in geometrized units with $G=c=1$. Greek indices $\mu,\nu,\ldots$ denote spacetime components, whereas Latin indices $i,j,\ldots$ refer to spatial ones. The metric signature is $(-+++)$. A dot above a variable denotes differentiation with respect to coordinate time $t$.

\section{Theoretical foundations}\label{sec: Section2}
    \subsection{Metric}
        The spacetime considered in this work is described by a perturbed Schwarzschild geometry (see Ref.~\cite{Polcar2022}), which incorporates the influence of a distant rotating mass ring. The ring, with mass $\mathscr{M}_r$, is located at a radius $r_r \gg M$. In Schwarzschild-like coordinates $(t,r,\theta,\phi)$, and valid in the near region $r \ll r_{r}$, the line element takes the form:
        \begin{equation}\label{eq: Metric}
        \begin{split}
            {ds}^2_{r \ll r_r} = 
            & -\left(1-\frac{2M}{r}\right)\left(1+2\nu_\mathscr{Q}\right)dt^2 + \frac{1+2\chi_\mathscr{Q}-2\nu_\mathscr{Q}}{1-2M/r}dr^2 \\
            & + (1-2\nu_\mathscr{Q})r^2\left[(1+2\chi_\mathscr{Q})d\theta^2+\sin^2\theta d\phi^2\right] ,
        \end{split}
        \end{equation}
        with the perturbation functions
        \begin{subequations}\label{eq: Metric parts}
        \begin{alignat}{4}
            \nu_\mathscr{Q} &\equiv \frac{\mathscr{Q}}{4}\left[r(2M-r)\sin^2\theta+2(M-r)^2\cos^2\theta-6M^2\right] , \\
            \chi_\mathscr{Q} &\equiv \mathscr{Q} M(M-r)\sin^2\theta ,
        \end{alignat}
        \end{subequations}
        where $\mathscr{Q}$ is quadrupole parameter defined as $\mathscr{Q} \equiv \mathscr{M}_r/r_r^3$. Compared with the unperturbed Schwarzschild solution, the spherical symmetry is broken by the presence of the ring, reducing the spacetime to an axisymmetric one. This metric serves as the dynamical background for all the subsequent analysis of particle motion and gravitational-wave fluxes in this article.
        
    \subsection{Integrability and non-integrability} \label{sec:dynamics}
        The axisymmetric spacetime~\eqref{eq: Metric} possesses two Killing vector fields, associated with the coordinate time $t$ and the azimuthal angle $\phi$. These fields give rise to two conserved quantities: the total energy $E$ and the axial component of angular momentum $L_z$, both defined with respect to an observer at infinity. Together with the third conserved quantity, the rest mass of the secondary $\mu$, these integrals of motion are insufficient for system to be integrable according to Liouville's theorem of integrability (see Ref.~\cite{Iro2016}). Nevertheless, the existing symmetries allow us to reduce our system to a two degrees of freedom (DoF) problem, described solely by coordinates $r$ and $\theta$. The reduced Hamiltonian then takes the form:
        \begin{equation}\label{eq: Reduced Hamiltonian}
            \mathcal{H} = \frac{1}{2} \left( \frac{\left( p_r \right)^2}{g_{rr}}  + \frac{\left( p_\theta \right)^2}{g_{\theta \theta}} + \frac{E^2}{g_{tt}} + \frac{L_z^2}{g_{\phi \phi}} \right) = - \frac{1}{2}\mu^2.
        \end{equation}
        
        For an integrable system, the Liouville-Arnold theorem \citep{Arnold1989} implies that, in a Hamiltonian system with two DoF, the bounded motion is confined to a two-dimensional invariant torus characterized by two fundamental frequencies, $\omega^1$ and $\omega^2$. The nature of the motion can be distinguished by the ratio $\omega = \omega^1 / \omega^2$. If $\omega$ is irrational, the motion is quasiperiodic: the trajectory densely covers the torus over infinite time and never returns to its initial state within a finite period. If $\omega$ is rational, the torus is said to be resonant, and the motion is periodic.
        
        When an integrable system is perturbed, as in the studied spacetime, the transition from integrability to non-integrability is governed by two fundamental theorems: the Kolmogorov-Arnold-Moser (KAM) theorem and the Poincaré-Birkhoff theorem. The KAM theorem asserts that non-resonant tori survive under sufficiently small perturbations~\cite{Iro2016}. Such invariant structures are referred to as KAM tori. Furthermore, the Poincaré-Birkhoff theorem states that at spaces where resonant tori previously existed, an even number of periodic trajectories persist, half of which are stable and half unstable, forming so-called Birkhoff chains~\cite{Poincare1912,Birkhoff1913}.

        \begin{figure}[t]\centering
        \includegraphics[width=0.48\textwidth]{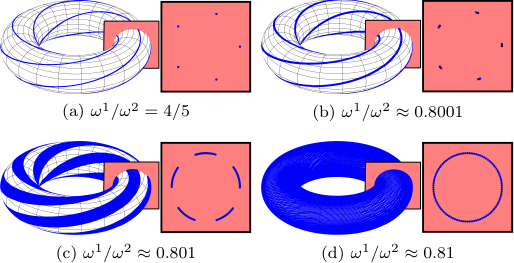}
        \caption{
        Illustration of motion on resonant (a) and quasiperiodic (b-d) tori in a two DoF system. Blue curves denote trajectories on the torus, while the Poincaré surfaces of section are in red. Each trajectory is evolved until it intersected the section 100 times. The resonant trajectory (a) corresponds to the resonant ratio $\omega^{1}/\omega^{2} = 4/5$. In contrast, the quasiperiodic trajectories (b-d) correspond to irrational ratios, yielding closed invariant curves that densely fill the torus over time.}
        \label{fig: Torus with Poincare sections}
        \end{figure}

        \begin{figure}[t]\centering
        \includegraphics[width=0.48\textwidth]{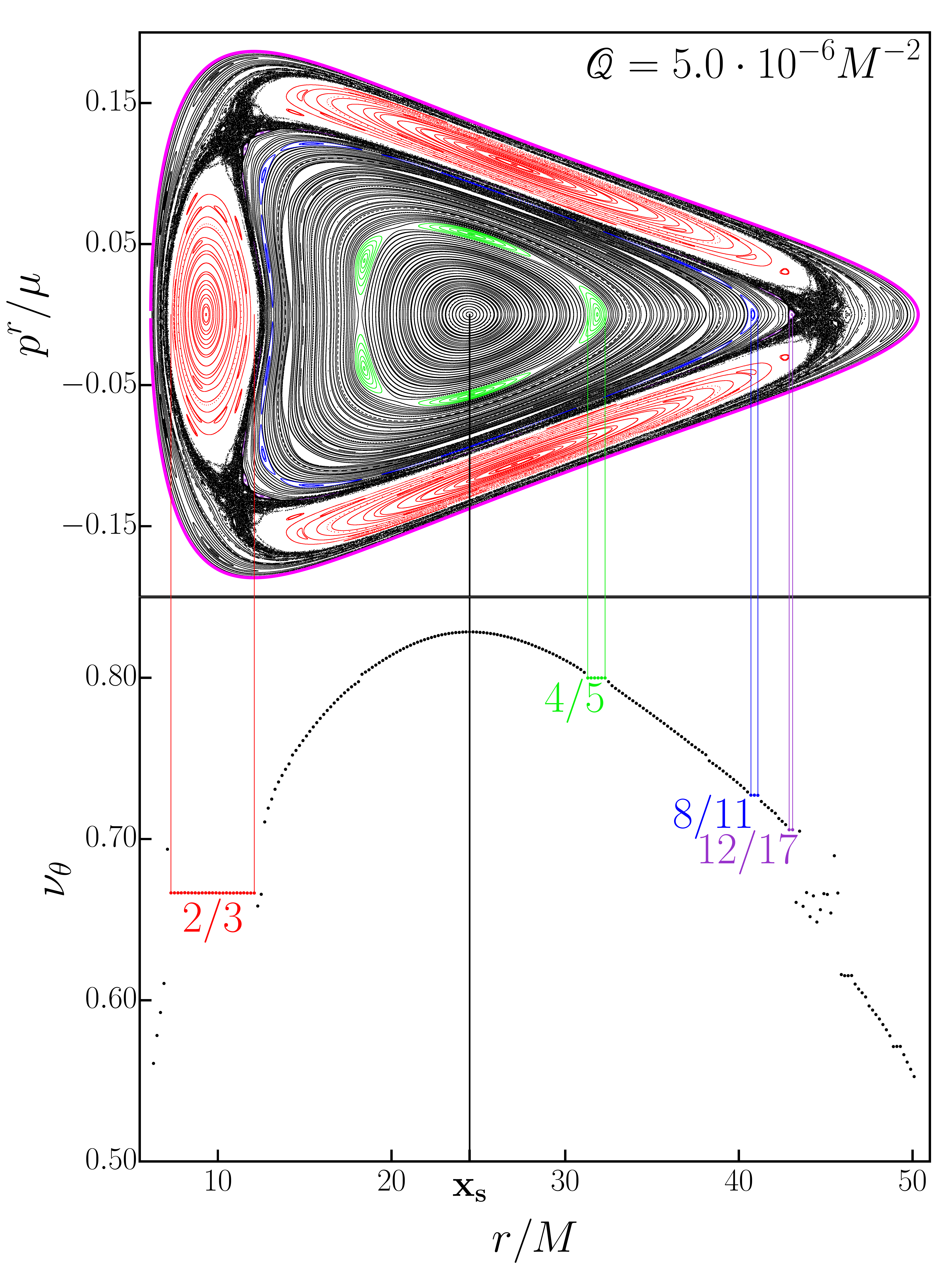}
        \caption{
        Top: Poincaré surface of section. Bottom: rotation curve along the $p^{r}/\mu=0$ line, with dominant resonances labeled by their frequency ratios. Parameters: ${L_z=4.0\mu M}$, ${E=0.98 \mu}$, ${\theta \left[ 0 \right]= \pi/2}$ and ${r \left[ 0 \right] \in \left( 6.298M; 50.098M \right)}$ with step size $0.2M$.}
        \label{fig: Poincare_Rotation}
        \end{figure}
        
        To analyze the dynamics of a Hamiltonian system with two DoF, one employs its symplectic structure to reduce the four-dimensional phase space to a two-dimensional slice, known as the \textit{Poincaré surface of section}. This construction requires identifying a section through the foliation of invariant tori such that the Hamiltonian flow intersects it transversely (see, e.g., Ref.~\cite{Lukes_Gerakopoulos2012}). Fig.~\ref{fig: Torus with Poincare sections} illustrates four trajectories and their corresponding Poincaré sections. In these sections, quasiperiodic trajectories form a sequence of nested, non-intersecting closed curves, whereas the resonant trajectories form a finite set of periodically repeated points.
        
        Another useful tool is the technique of \textit{rotation numbers}~\cite{Voglis1990,Voglis1998}, which assigns to each trajectory a specific rotation number, defined as:
        \begin{equation}\label{eq: Rotation number}
            \nu_\vartheta = \lim_{N\to\infty} \frac{1}{2\pi N}\sum^N_{i=1}\vartheta_i,
        \end{equation}
        where $\vartheta_i$ denotes the angle, measured with respect to the center of the main island of stability $\mathbf{x_s}$ (see Fig.~\ref{fig: Poincare_Rotation}), between two consecutive intersections of the trajectory with the Poincaré section. In the limit $N\to\infty$, this rotation number converges to the ratio $\omega$, with error bounded by $1/N$ for finite $N$~\cite{Voglis1998}.
        
        By plotting the rotation number as a function of the initial radial distance from center of the tori foliation, one obtains the \textit{rotation curve}. In integrable non-degenerate systems, this curve is strictly monotonic as one advances away from the center of the main island of stability $\mathbf{x}_s$. In contrast, in perturbed systems near the resonances, the rotation curve starts to fluctuate randomly due to emergence of a chaotic layer. Furthermore, within the Birkhoff chains, stable regions known as islands of stability appear as plateaus of constant value in the rotation curve (see bottom panel of Fig.~\ref{fig: Poincare_Rotation}).
        
    \subsection{Quadrupole formula}
        One of the most significant results of the linearized theory of gravity is the quadrupole formula, which provides a leading-order description of gravitational radiation from a non-relativistic source. In this framework, the metric is treated as a small perturbation of flat spacetime, and the dominant contribution to gravitational-wave emission arises from the time variation of the traceless mass quadrupole moment of the system.
        
        The particle's traceless quadrupole moment is defined as~\cite{Maggiore2007}:
        \begin{equation}
            Q_{ij}(t) = \mu \bigg( x_{i} (t) x_{j} (t) - \tfrac{1}{3}\delta_{ij} x^k (t) x_k (t) \bigg) .
        \end{equation}
        Note that functions $x^i(t)$ represent the Cartesian coordinates. Since the dynamics are evolved in Schwarzschild-like coordinates, it is necessary to use a standard
        flat-space transformation.
        
        The average energy flux radiated in gravitational waves is then given by~\cite{Maggiore2007}:
        \begin{equation}
            \left\langle \frac{dE}{dt} \right\rangle_{\mathrm{QP}} = \frac{1}{5} \big\langle \dddot{Q}_{ij}\dddot{Q}^{ij}\big\rangle ,
        \end{equation}
        while the flux of angular momentum reads
        \begin{equation}
            \left\langle \frac{dL^{i}}{dt} \right\rangle_{\mathrm{QP}} = \frac{2}{5} \, \epsilon^{ijk} \big\langle \ddot{Q}_{jl}\dddot{Q}_{kl} \big\rangle .
        \end{equation}
        
        The derivatives are with respect to the coordinate time $t$ and can be expressed in terms of proper time $\tau$ using~\cite{Polcar2022}:
        \begin{equation}
            \frac{d^n}{dt^n} \approx \left( \frac{d \tau}{dt} \right)^n \frac{d^n}{d \tau^n}.
        \end{equation}
        
        These quadrupole (QP) fluxes, first derived by Einstein, are widely used in gravitational-wave physics, particularly in weak-field regimes where higher multipole contributions are subdominant. Although approximate, the quadrupole formula remains a cornerstone of gravitational-wave modeling and provides an essential reference point for comparison with more sophisticated approaches such as post-Newtonian expansions or Teukolsky-based computations.
        
    \subsection{Post-Newtonian formalism}
        In the weak-field regime and slow motion, gravitational-wave emission can be described using the post-Newtonian (PN) approximation, which systematically expands the fluxes in powers of the orbital velocity. They are, therefore, well suited for the description of systems in which the gravitational field is relatively weak and the orbital velocities are small compared to the speed of light. The key advantage of these PN expressions is their closed analytic form, which allows rapid evaluation and integration over long timescales, making them highly useful in semi-analytic inspiral modeling.
        
        For a bound geodesic in Kerr spacetime, specified by semilatus rectum $p$, eccentricity $e$, and inclination $\iota$ (for definition see Ref.~\cite{Glampedakis2002}), the averaged lowest order post-Newtonian fluxes of energy and angular momentum are given by \cite{Gair2006}:
        
        \begin{widetext}
        \begin{equation}\label{eq: PN fluxes}
        \begin{aligned}
            \left( \frac{dE}{dt} \right)_{\text{PN}} &= \frac{32}{5} \frac{\mu^2}{M^2} \left( \frac{M}{p} \right)^5 \Big(1 - e^2\Big)^{3/2} \left[  f_1 - \frac{a}{M} \left( \frac{M}{p} \right)^{3/2} \cos \iota f_2 \right], \\
            \left( \frac{dL_z}{dt} \right)_{\text{PN}} &= \frac{32}{5} \frac{\mu^2}{M} \left( \frac{M}{p} \right)^{7/2} \Big(1 - e^2\Big)^{3/2} \left[ \cos \iota f_3 + \frac{a}{M} \left( \frac{M}{p} \right)^{3/2} \left( f_4 - \cos^2 \iota f_5 \right) \right],
        \end{aligned}
        \end{equation}
        \end{widetext}
        where $M$ is the black hole mass, $\mu$ the secondary mass, $a$ the spin parameter, and the eccentricity-dependent coefficients are defined as follows:
        \begin{align*}
            f_1 &= 1 + \frac{73}{24} e^2 + \frac{37}{96} e^4, \\
            f_2 &= \frac{73}{12} + \frac{823}{24} e^2 + \frac{949}{32} e^4 + \frac{491}{192} e^6, \\
            f_3 &= 1 + \frac{7}{8} e^2, \\
            f_4 &= \frac{61}{24} + \frac{63}{8} e^2 + \frac{95}{64} e^4, \\
            f_5 &= \frac{61}{8} + \frac{91}{4} e^2 + \frac{461}{64} e^4.
        \end{align*}
        
        In the non-spinning limit $a \to 0$, these expressions simplify considerably, and the energy flux reduces to the well-known Peters–Mathews formula~\cite{Peters1963}. Note that also definition of inclination reduces to its usual interpretation. Although limited to integrable motion with constant inclination, the PN approximation provides compact analytic expressions that are valuable for benchmarking more sophisticated numerical approaches.
        
    \subsection{Teukolsky equation}
        As our work concerns motion around a black hole, a more accurate description of gravitational radiation is provided by black hole perturbation theory. In this framework, gravitational perturbations of a Kerr (or Schwarzschild) spacetime can be encoded in the Weyl scalar $\Psi_{4}$, which satisfies the Teukolsky master equation (see Ref.~\cite{Teukolsky1973}). This formalism is based on the Newman-Penrose tetrad approach and yields a gauge-invariant wave equation for $\Psi_{4}$ with spin weight $s=-2$. In the Schwarzschild limit ($a=0$), the master equation reduces to form:
        \begin{widetext}
        \begin{equation}\label{eq: Teukolsky equation}
        \begin{aligned}
            \frac{r^2}{f} \frac{\partial^2 \psi}{\partial t^2}
            &- \frac{1}{\sin^2 \theta} \frac{\partial^2 \psi}{\partial \phi^2}
            - \left( \frac{1}{r^2 f} \right)^s \frac{\partial}{\partial r} \left[ (r^2 f)^{s+1} \frac{\partial \psi}{\partial r} \right]
            - \frac{1}{\sin \theta} \frac{\partial}{\partial \theta} \left( \sin \theta \frac{\partial \psi}{\partial \theta} \right) \\
            &- 2s \frac{i \cos \theta}{\sin^2 \theta} \frac{\partial \psi}{\partial \phi}
            - 2s \left( \frac{M}{f} - r \right) \frac{\partial \psi}{\partial t}
            + s \left( s \cot^2 \theta - 1 \right) \psi
            = 4 \pi r^2 T,
        \end{aligned}
        \end{equation}
        \end{widetext}
        with master variable $\psi = r^{-4} \Psi_4$ and source term $T$ (see Ref.~\cite{Teukolsky1973}). Moreover, $f$ is Schwarzschild function given as $f(r) = 1 - 2M/r$.
        
        The Teukolsky equation (TKEQ) is a second-order partial differential equation in the time domain. However, in an axisymmetric spacetime, the problem reduces to a $(2+1)$-dimensional system in the variables $(t, r, \theta)$, and admits the decomposition~\cite{Harms2014}:
        \begin{equation}\label{eq: Decomposition into m-modes}
            \psi (t,r,\theta,\phi) = \sum_{m = -\infty}^\infty e^{im\phi} \psi_m (t,r,\theta).
        \end{equation}
        Because the master equation is linear, the different $m$-modes decouple, allowing Eq.~\eqref{eq: Teukolsky equation} to be split into independent sub-equations for each $\psi_m(t,r,\theta)$.
        
        Numerical solutions of this system enable extraction of the gravitational-wave strain $h_m$ at future null infinity (see Ref.~\cite{Harms2014}). From that, one computes the fluxes of energy and angular momentum. In particular, the Teukolsky formalism yields
        \begin{equation}\label{eq: Teukolsky fluxes}
        \begin{aligned}
            \left( \frac{dE}{dt} \right)_{\text{TK}} &= \frac{1}{16 \pi} \sum_m \int^\pi_0 \left| r \accentset{\mbox{\large\bfseries .}}{h}_m \right|^2 \sin{\theta} d\theta , \\
            \left( \frac{dL_z}{dt} \right)_{\text{TK}} &= \frac{1}{16 \pi} \mathcal{I} \left\{ \sum_m m \int^\pi_0 \left( \overline {r \accentset{\mbox{\large\bfseries .}}{h}_m} \right) \left( r h_m \right) \sin{\theta} d\theta \right\} ,
        \end{aligned}
        \end{equation}
        
        Among the three approaches discussed in this work, the Teukolsky formalism is the most advanced, but also the most computationally demanding, as it requires, in general, sophisticated numerical solvers. In this study, we employ the time-domain code \textit{Teukode}, which implements the Teukolsky formalism in hyperboloidal, horizon-penetrating (HH) coordinates, enabling direct wave extraction at null infinity~\cite{Harms2014}.
  
\section{Comparison of the fluxes}\label{sec: Section3}
    \subsection{Numerical setup}
        We compare the performance of the three methods described in the previous section for computing GW fluxes under a variety of initial conditions. The orbital dynamics are evolved by numerically integrating Hamilton's equations based on the reduced Hamiltonian~\eqref{eq: Reduced Hamiltonian} with a standard fourth-order Runge-Kutta algorithm implemented in C.
        
        In the QP formalism, the complexity of higher-order derivatives of the quadrupole moment grows rapidly. To maintain computational efficiency, we evaluated the first derivative analytically, while the second and third derivatives were computed using finite-difference schemes with fourth-order accuracy.
        
        For the TK method, we computed the first six azimuthal modes ($m=0$–$5$), which together account for at least $95\%$ of the total flux contribution~\cite{Harms2016}. Due to the axial and reflection symmetries, it is sufficient to compute only positive $m$ modes and double their contribution to account for negative $m$ values. As expected from Eq.~\eqref{eq: Teukolsky fluxes}, the $m=0$ mode does not contribute to angular momentum fluxes.
        
        To ensure consistent averaging of fluxes across different orbital configurations, we introduced a \textit{tolerance test}. This procedure determines the endpoint of orbital evolution by combining two conditions: a prescribed minimum number of revolutions $n_{min}$ of the secondary object around the primary and a tolerance threshold $\mathcal{T}$ on the radial return distance. For more details about these auxiliary quantities, we refer the interested reader to Appendix~\ref{sec: Tolerance test}. The method guarantees that averaged fluxes are computed over dynamically consistent segments of the trajectory, even in non-periodic perturbed cases.
        
        As a baseline, we first validated the proposed methods in the unperturbed Schwarzschild spacetime to verify the basic behavior of the methods; detailed results are presented in Appendix~\ref{sec:EquatFlux}. We then moved to the perturbed Schwarzschild metric~\eqref{eq: Metric}, which we analyze in the next section.
        
    \subsection{Perturbed motion}   
        Let us now examine the fluxes from the motion in a perturbed Schwarzschild spacetime, described by Eq.~\eqref{eq: Metric}. In this spacetime, none of the methods are formally exact: the PN approximation assumes constant orbital parameters and breaks down entirely once eccentricity and inclination cease to be constants, while both the quadrupole QP and TK methods provide only approximate fluxes in a nearly integrable system, i.e., when the perturbation is reasonably small. Another issue that arises when we depart from integrability is the non-ergodic coverage of the phase space by the trajectory near the stable periodic points of a Birkhoff chain, but we ignore this handicap in our study.
        
        Moreover, recall that the studied spacetime is not asymptotically flat, and therefore the computed fluxes via the QP and TK methods do not represent the actual fluxes at infinity. Note that, more rigorously, this problem has been tackled by considering an additional rotation of the spacetime~\cite{Polcar2022}. Nonetheless, a comparison between the fluxes remains instructive by providing qualitative insight.
        
        The initial parameters were chosen as $E = 0.98\mu$, ${L_z = 4.00\mu M}$, $r[0] = 20M$, $p^r[0] = 0$, $\theta[0] = \pi/2$. They were varied only in the quadrupole perturbation parameter $\mathscr{Q}$, spanning from effectively vanishing values ($10^{-16}M^{-2}$, numerically indistinguishable from Schwarzschild) up to strongly perturbed configurations ($10^{-5}M^{-2}$), where large chaotic layers and multiple resonances appear. The initial parameters were chosen to be sufficiently far from the resonances that develop at higher values of $\mathscr{Q}$, in order to study only the effect of the perturbation on a regular KAM trajectory (see Fig.~\ref{fig: Poincare_Rotation}).
        
        The Teukode simulations were carried out on a grid with $1001 \times 76$ points. The tolerance test was enabled with $\mathcal{T} = 0.005M$, $n_{\min} = 10$. In practice, this led to trajectory lengths varying between 10 and 55 revolutions.
        
        The modal decomposition shown in Fig.~\ref{fig: Perturbed motion - m-modes} confirms the unperturbed tests (see Appendix~\ref{sec:EquatFlux}) that the ${m = \pm 2}$ modes are dominant, while $m = \pm 1$ contributions are subleading and becoming more important as the perturbation increases. However, together, the modes $m = \pm 1, \pm 2$ account for more than $90\%$ of the fluxes across the entire range of $\mathscr{Q}$.

        \begin{figure*}[t]\centering
        \includegraphics[width=1.0\textwidth]{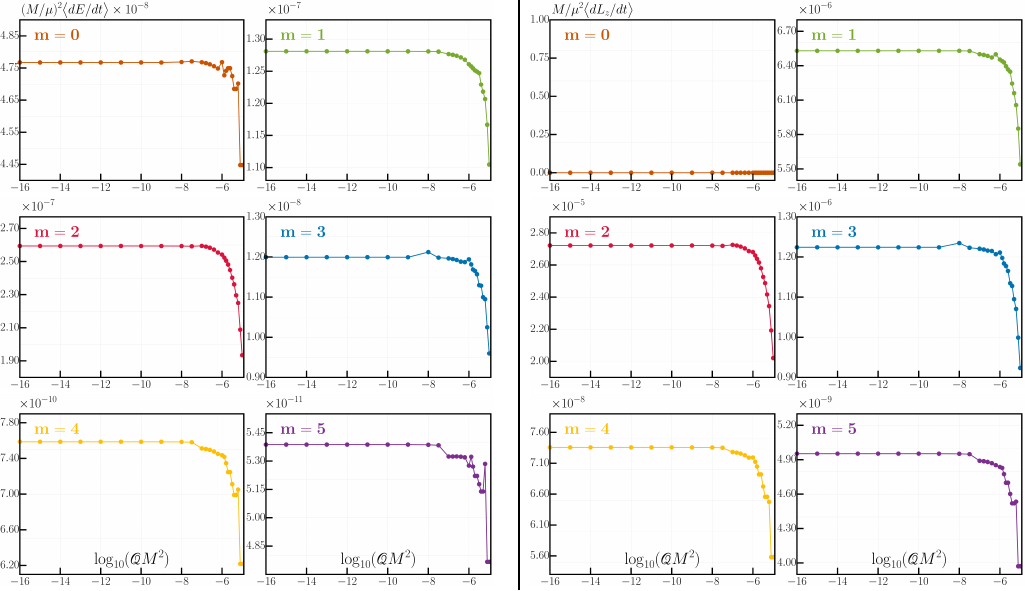}
        \caption{
        Individual TK $m$ mode contributions to the energy (left two columns) and angular momentum (right two columns) fluxes for the motion in the perturbed Schwarzschild spacetime~\eqref{eq: Metric} as a function of the logarithm of the quadrupole perturbation parameter.}
        \label{fig: Perturbed motion - m-modes}
        \end{figure*}

        \begin{figure*}[t]\centering
        \includegraphics[width=1.0\textwidth]{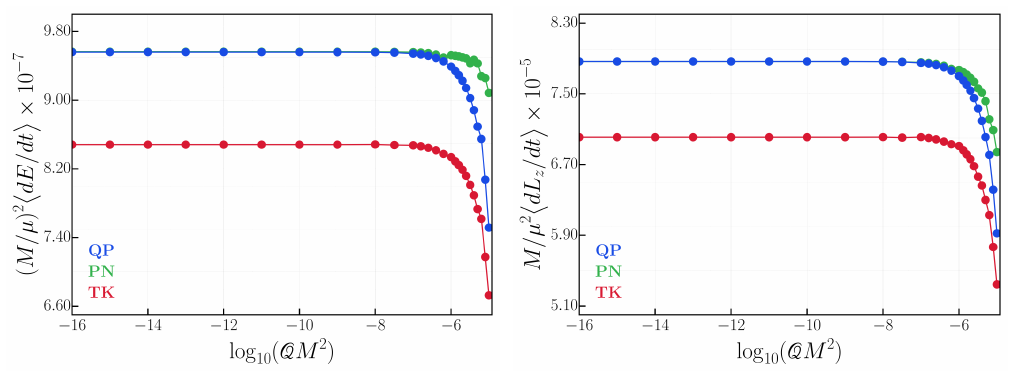}
        \caption{
        Comparison of energy (left) and angular momentum (right) fluxes computed using the three different methods for the motion in the perturbed Schwarzschild spacetime~\eqref{eq: Metric}.}
        \label{fig: Perturbed motion - full plot}
        \end{figure*}
        
        A comparison of the three methods is shown in Fig.~\ref{fig: Perturbed motion - full plot}. As expected, the PN approach diverges from the other two as the perturbation increases, reflecting its limited applicability outside of integrable spacetimes. By contrast, the QP and TK results remain consistent in both qualitative trends and absolute magnitudes, reinforcing their reliability even in moderately perturbed regimes.
        
        This agreement provides the basis for the study of orbital resonances presented in the next section, where the system enters non-integrable domains and resonance plateaus appear in the flux profiles.
        
        \begin{figure*}[t]\centering
        \includegraphics[width=1.0\textwidth]{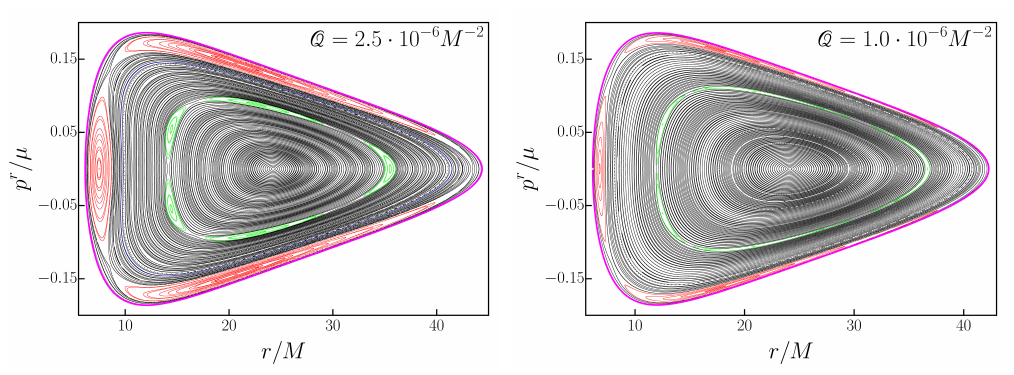}
        \caption{
        Poincaré surfaces of section for the spacetimes used in the study of the 2/3 resonance (left) and in the subsequent resonance crossing through 2/3 resonance within adiabatic inspiral (right). The parameters are identical to those described in the text. Initial radii were sampled with step size $0.2M$ over the intervals $(6.266M,44.266M)$ (left) and $(6.298M,50.098M)$ (right).
        }
        \label{fig: Poincare sections}
        \end{figure*}
        
    \subsection{Resonant motion}\label{sec: Resonant motion}
        In this section, we investigate orbital resonances in the perturbed Schwarzschild spacetime given by Eq.~\eqref{eq: Metric}. Since the PN approach breaks down in this non-integrable regime, we restrict our analysis to the QP and TK methods, which remain approximately valid in the near-integrable domain. The following subsections present detailed results for the two strongest resonances observed, namely the $2/3$ and $4/5$ cases.
        
    \subsubsection*{2/3 Resonance}
        The 2/3 resonance appears prominently in the perturbed Schwarzschild spacetime (see Fig.~\ref{fig: Poincare_Rotation}). In the chosen configuration $\mathscr{Q} = 2.5 \cdot 10^{-6}M^{-2}$, the resonance is sufficiently large, while the surrounding chaotic layer is negligible (see Fig.~\ref{fig: Poincare sections}), allowing a clean scan across the resonance structure and its impact on the GW fluxes.
        
        The parameters were set to $E = 0.98\mu$, $L_z = 4.00\mu M$, $p^r[0] = 0$, $\theta[0] = \pi/2$, $\dot{\theta}[0] = 0$. Initial conditions were sampled with $r[0] \in (6.2M, 9.0M)$, step $\Delta r = 0.1M$, and finer resolution ($\Delta r = 0.02M$) near the resonance edges. The Teukode simulations used a $801 \times 61$ grid, with the tolerance test's parameters set to $\mathcal{T}=0.001M$ and $n_{\min}=25$.
        
\begin{figure*}[t]\centering
        \includegraphics[width=1.0\textwidth]{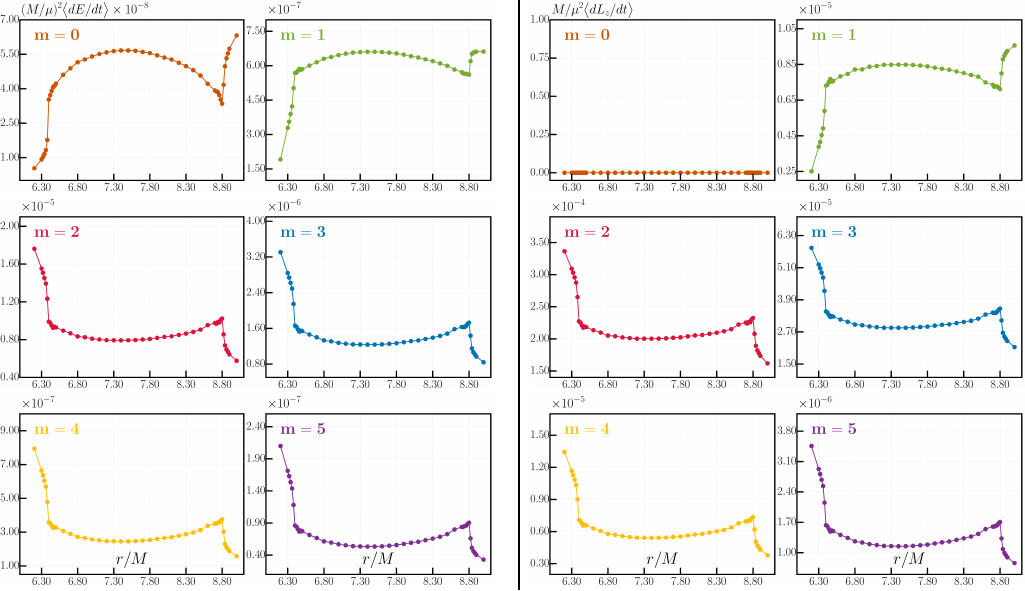}
        \caption{Individual TK $m$ mode contributions to the energy (left two columns) and angular momentum (right two columns) fluxes for the motion in the vicinity of the $2/3$ resonance in the perturbed Schwarzschild spacetime~\eqref{eq: Metric}.}
        \label{fig: Resonance 2/3 - m-modes}
        \end{figure*}

        The modal decomposition of TK fluxes shown in Fig.~\ref{fig: Resonance 2/3 - m-modes} confirms that the $m=\pm 2$ modes remain dominant. Moreover, with the subleading modes $m=\pm 3$, they together contribute $\gtrsim 88\%$ to the total flux throughout the resonance width. We observe a characteristic plateau (or rather $U$-shaped plateau) structure with asymmetry in the modal contributions when comparing the left and right sides next to the resonance.
        
\begin{figure}[t]\centering
        \includegraphics[width=0.48\textwidth]{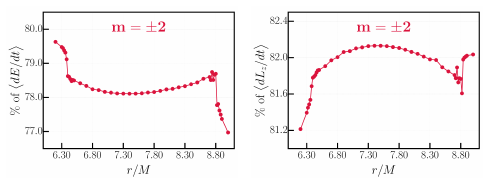}
        \caption{
        Percentage contribution of $m = \pm 2$ mode to the total energy flux (left) and to the total angular momentum flux (right) for motion in the vicinity of the $2/3$ resonance in the perturbed spacetime~\eqref{eq: Metric}.
        }
        \label{fig: Contribution of m = 2}
        \end{figure}

        Moreover, as the initial radius approaches the main island of stability, the relative contribution of $m = \pm 2$ to energy fluxes decreases, while the contribution to angular momentum fluxes increases (see Fig.~\ref{fig: Contribution of m = 2}). This inverse trend between energy and angular momentum fluxes is present in other modes as well.
        
\begin{figure*}[t]\centering
        \includegraphics[width=1.0\textwidth]{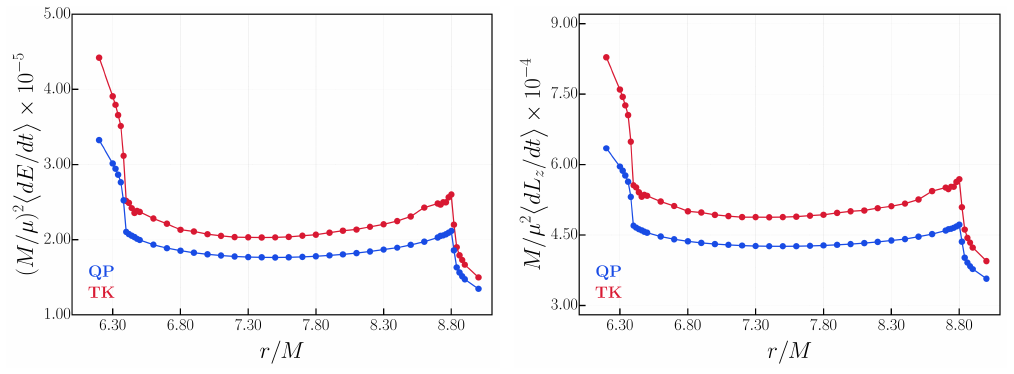}
        \caption{Comparison of energy (left) and angular momentum (right) fluxes computed using the two different methods for the motion in the vicinity of the $2/3$ resonance in the perturbed Schwarzschild spacetime~\eqref{eq: Metric}.}
        \label{fig: Resonance 2/3 - full plot}
        \end{figure*}

        Figure~\ref{fig: Resonance 2/3 - full plot} compares the averaged energy and angular momentum fluxes computed using the QP and TK methods. Both exhibit the expected $U$-plateau structure across the resonance, with good agreement in both trend and magnitude. The $U$-plateaus reflects clearly their symmetry about the resonance center.
        
    \subsubsection*{4/5 Resonance}
        The 4/5 resonance lies in a more complex phase-space structure (see Fig.~\ref{fig: Poincare_Rotation}). Namely, a noticeable chaotic layer surrounds the islands of stability, but this layer still allows a clean comparison of fluxes within the 4/5 resonance.
        
        The parameters were chosen as $\mathscr{Q} = 5.0 \cdot 10^{-6}M^{-2}$, $E = 0.98\mu$, $L_z = 4.00 \mu M$, $p^r[0] = 0$, $\theta[0] = \pi/2$, ${\dot{\theta}[0] = 0}$. The initial radii $r[0] \in (30.95M, 32.60M)$ range is covered with step $\Delta r = 0.05M$, and it is refined to $\Delta r = 0.01M$ near the resonance edges. The Teukode computations used a $801 \times 61$ grid, with the tolerance test's parameters set to $\mathcal{T} = 0.005M$ and $n_{\min} = 30$.
        
\begin{figure*}[t]\centering
        \includegraphics[width=1.0\textwidth]{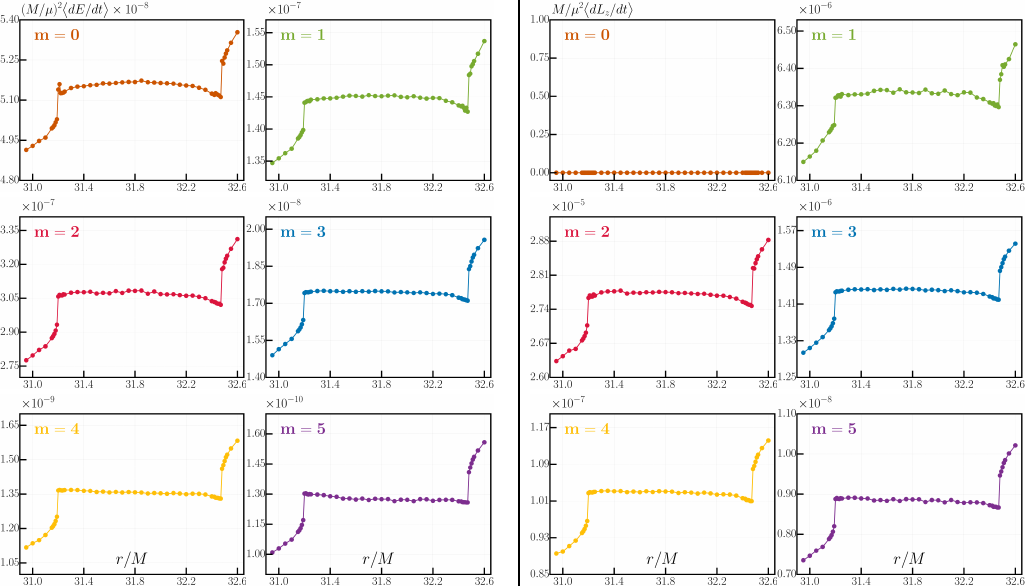}
        \caption{Individual TK $m$ mode contributions to the energy (left two columns) and angular momentum (right two columns) fluxes for the motion in the vicinity of the $4/5$ resonance in the perturbed Schwarzschild spacetime~\eqref{eq: Metric}.}
        \label{fig: Resonance 4/5 - m-modes}
        \end{figure*}

        The decomposition of TK fluxes is shown in Fig.~\ref{fig: Resonance 4/5 - m-modes} and again identifies the $m = \pm 2$ modes as dominant, but here the subleading contribution shifts to $m = \pm 1$. Together, these modes carry over $90\%$ of the total flux across the full resonance width. Unlike the 2/3 resonance, the modal contributions to energy and angular momentum fluxes vary in the same direction with increasing distance from the main island of stability, highlighting the distinct dynamical character of this resonance.

        \begin{figure*}[t]\centering
        \includegraphics[width=1.0\textwidth]{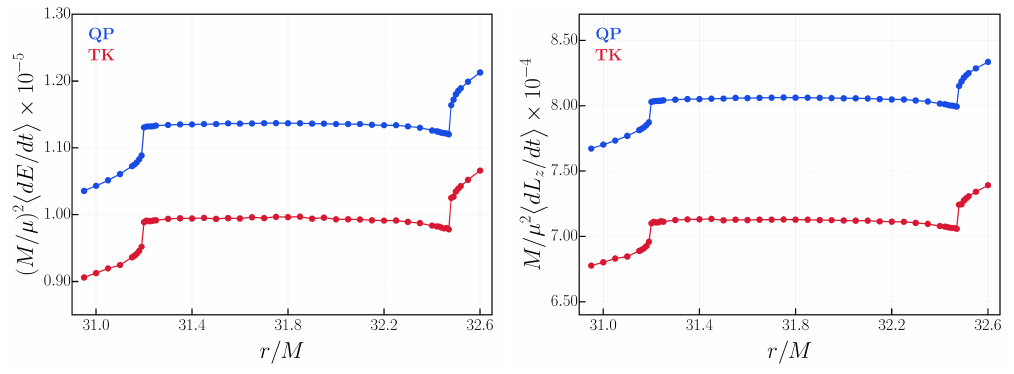}
        \caption{Comparison of energy (left) and angular momentum (right) fluxes computed using the two different methods for the motion in the vicinity of the $4/5$ resonance in the perturbed Schwarzschild spacetime~\eqref{eq: Metric}.}
        \label{fig: Resonance 4/5 - full plot}
        \end{figure*}
        
        In Fig.~\ref{fig: Resonance 4/5 - full plot}, the QP and TK results are compared directly. Both methods exhibit the characteristic plateau structure of resonant fluxes and remain consistent in both magnitude and qualitative behavior.
        
        These results for both resonances are consistent with independent estimates of resonance width obtained via rotation curves in earlier work~\cite{Strateny2023a}, confirming that the flux-based approach captures the resonant dynamics. The close agreement between QP and TK methods indicate that the computed fluxes can be used to study resonance crossings during inspirals in non-integrable spacetime.

\section{Adiabatic inspiral}\label{sec: Section4}
        The concept of adiabatic inspiral exploits the two time-scale separation in an EMRI \cite{Hinderer2008}: while the orbital motion around the central black hole occurs on dynamical timescales $\sim M$, the inspiral, which is caused by the radiation-reaction, acts on a much longer timescale $\sim M/\mu = q^{-1}$. In this approximation, the inspiral of a non-spinning compact object into a supermassive black hole is modeled as a sequence of quasi-stationary geodesics. In the absence of gravitational radiation, the orbit would be an exact geodesic on the background spacetime, parametrized by three conserved quantities $(E,L_z,Q)$, where $Q$ is the Carter constant. When GW emission is included, these quantities evolve slowly due to the continual loss of energy and angular momentum~\cite{Glampedakis2002,Hughes2021}.

        In asymptotically flat spacetimes, the rates of change of the conserved quantities are determined by flux-balance laws, which equate the losses of energy and angular momentum to the fluxes carried away to null infinity and down the black hole horizon~\cite{Isoyama2019}. The orbital worldline is, thus, reconstructed by evolving the constants of motion while assuming that, locally, the trajectory remains geodesic.
        
        In a perturbed, non-integrable spacetime, where no Carter-like constant exists, the adiabatic evolution is restricted to $E$ and $L_z$ fluxes \cite{Gair2008,Destounis2021,Lukes-Gerakopoulos2022,Destounis2023,Mukherjee2023,Destounis2025}. The dissipative dynamics is then given by~\cite{Isoyama2019,Polcar2022}:
        \begin{equation}
            \frac{dE}{dt} = -\langle \dot{E} \rangle, 
            \qquad 
            \frac{dL_z}{dt} = -\langle \dot{L}_z \rangle ,
        \end{equation}
        where angle brackets denote appropriate time-averaging of the GW fluxes over many orbital periods. Therefore, radiation-reaction corrections enter through the time evolution of $E$ and $L_z$ only. The remaining components of the four-momentum, namely $P^r$ and $P^\theta$, are advanced using the conservative geodesic equations, where the dissipative effects enter from the dependence on the dissipating $E$ and $L_z$.
        
        This adiabatic scheme provides a consistent and computationally manageable method for modeling EMRI dynamics, wherein the secondary's orbit is approximated by a continuous sequence of geodesic trajectories governed by the gradual orbital loss of energy and angular momentum driven by GW emission.

\section{Resonance crossing}\label{sec: Section5}
        Building on the flux comparison, we employ the QP formalism to model adiabatic inspirals. The procedure is straightforward: we evolve the trajectory by updating, at each integration step, the energy and axial angular momentum provided by the QP formalism. However, this direct approach is numerically very demanding. Therefore, to produce several hundred inspirals, we precompute GW fluxes on a grid in $(E, L_z, e)$ and evaluate the actual fluxes during the evolution via interpolation. We use the eccentricity-like parameter $e$ as a practical third coordinate (see Appendix~\ref{sec: Parametrization of the trajectory}) and employ a robust spline-based interpolant on the flux grid (see Appendix~\ref{sec: Interpolation}), thereby avoiding per-step flux direct computations.
        
        In addition, we introduce a \textit{repetitiveness parameter} $\mathcal{R}$ that updates the fluxes once every $\mathcal{R}$ integrator steps, exploiting their slow variation along the inspiral (see Appendix~\ref{sec: Repetitiveness parameter}). 
        
\begin{figure}[t]\centering
        \includegraphics[width=0.48\textwidth]{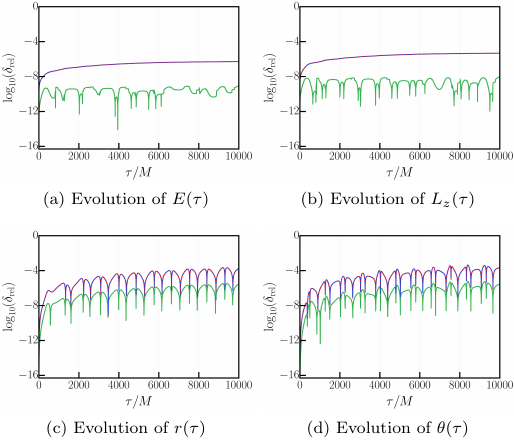}
        \caption{Final performance test comparing the evolution of orbital parameters. Relative errors are computed with respect to the reference case ($\mathcal{R}=1$, interpolation disabled). 
        Red: $\mathcal{R}=1$ with interpolation; green: $\mathcal{R}=100$ without interpolation; blue: $\mathcal{R}=100$ with interpolation. 
        Initial conditions: $\mathscr{Q}=5.0\times10^{-6} M^{-2}$, $E[0]=0.98\mu$, $L_z[0]=4.0\mu M$, $r[0]=32.4840M$, $p^r[0]=0$, $\theta[0]=\pi/2$, $\Delta\tau=0.25M$, $n=10$, $\mathcal{T}=0.01M$, $q=10^{-3}$.}
        \label{fig: Final test}
        \end{figure}

        Together, these methods reduce computational cost by orders of magnitude and enable systematic scans of hundreds of inspirals while maintaining acceptable accuracy. In a final benchmark, see Fig.~\ref{fig: Final test}, the repetitiveness parameter contributes errors that are several orders of magnitude below the interpolation error (the red and blue curves overlap), while the relative errors of $(E,L_z,r,\theta)$ remain safely below thresholds that would compromise adiabatic evolution. The wall-clock time for an integration interval of $\tau=10\,000M$ drops from $\sim 188$ minutes (no optimizations) to $\sim 36$ minutes with interpolation, to under 2 minutes with repetitiveness alone, and to $\sim 21$ seconds when both are combined. Therefore, this yields an overall $\sim 500\times$ speedup without a significant loss of accuracy.

        \begin{figure*}[t]\centering
        \includegraphics[width=1.0\textwidth]{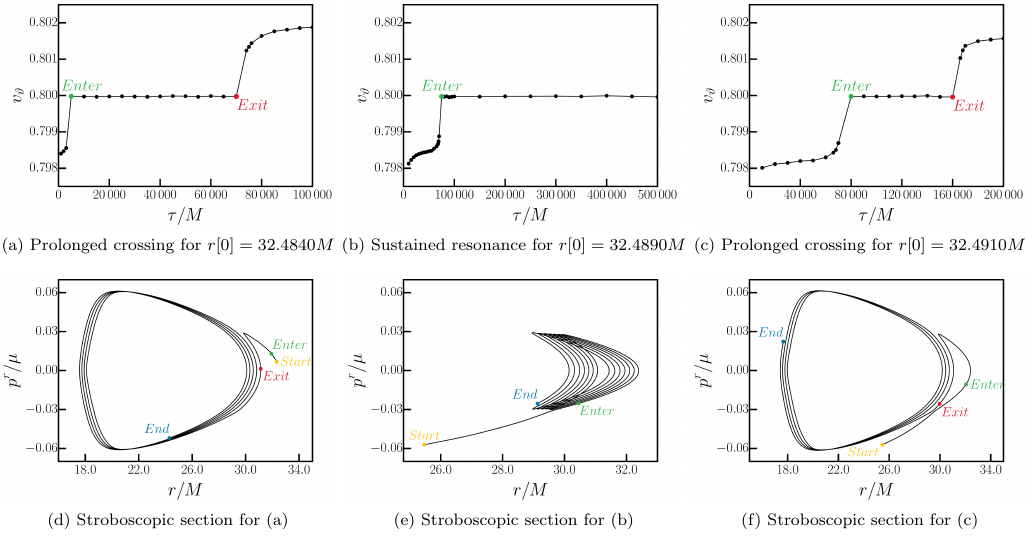}
        \caption{General resonance crossings through the 4/5 resonance. Top row: rotation number along the inspiral, evaluated by pausing dissipation. Bottom row: stroboscopic Poincaré sections with every fifth intersection plotted. Other initial parameters are $\mathscr{Q} = 5.0\cdot10^{-6} M^{-2}$, $E[0] = 0.98\mu$, $L_z[0] = 4.0\mu M$, $p^r[0] = 0$, $\theta[0] = \pi/2$, $\Delta \tau = 0.25M$, $\mathcal{R} = 200$ and mass ratio $q = 10^{-3}$. The Poincaré sections were evolved for $\tau = 10^6M$.}
        \label{fig: General resonance crossings through 4/5}
        \end{figure*}
        
    \subsection{General resonance crossing}
        
        \begin{figure}[t]\centering
        \includegraphics[width=0.48\textwidth]{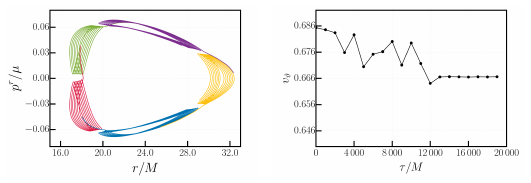}
        \caption{The left panel displays the entire Poincaré section from the resonant trapping case depicted in Fig.~\ref{fig: General resonance crossings through 4/5} divided into five distinct sets. The initial parameters are unchanged. The right panel shows different rotation curve for $\mathscr{Q} = 2.5\cdot10^{-6} M^{-2}$, $E[0] = 0.98 \mu$, $L_z[0] = 4.0\mu M$, $r[0] = 8.8254 M$, $p^r[0] = 0$, $\theta[0] = \pi/2$, $\Delta \tau = 0.25M$, $\mathcal{R} = 500$, and mass ratio $q = 10^{-3}$. The fluxes were calculated directly.}
        \label{fig: Additional plots}
        \end{figure}
        
        With the numerical infrastructure in place, we evolve adiabatic inspirals and characterize their passage through resonances in the perturbed Schwarzschild spacetime~\eqref{eq: Metric}.
        
        The initial data are chosen based on insights from Poincaré sections (see, e.g., Fig. ~\ref{fig: Poincare sections}). Due to the dissipation of energy and angular momentum, the orbit drifts between sections of this type. The resonance islands also shift during the inspiral. Therefore, to reliably observe a crossing, the starting position must be placed next to the appropriate side of the resonant island. In our setup, this side differs between the two resonances studied: for the 2/3 case, it is the side facing the main island, whereas for the 4/5 case, it is the opposite side. Otherwise, the inspiral proceeds without entering the resonance.
        
        Following Refs.~\cite{Mukherjee2023,Lukes-Gerakopoulos2022}, we diagnose a resonance with two complementary tools. First, we track the rotation number~\eqref{eq: Rotation number} along the inspiral. For a clean measurement, we periodically freeze the dissipative evolution and evaluate the rotation number in the corresponding conservative dynamics. A plateau at the rational value $P/Q$ indicates that the trajectory is crossing a resonance. Second, we construct Poincaré sections in a stroboscopic manner, plotting only every $Q$-th intersection for a $P/Q$ resonance. For example, at the 4/5 resonance, the section displays five islands; sampling every fifth intersection isolates the motion within a single island. This technique also applies to near-resonant trajectories that do not yet form resonance structures: selecting every fifth point reveals a smooth tracing of the corresponding KAM curve \cite{Lukes-Gerakopoulos2010,Lukes-Gerakopoulos2022}.
        
        \begin{figure*}[t]\centering
        \includegraphics[width=1.0\textwidth]{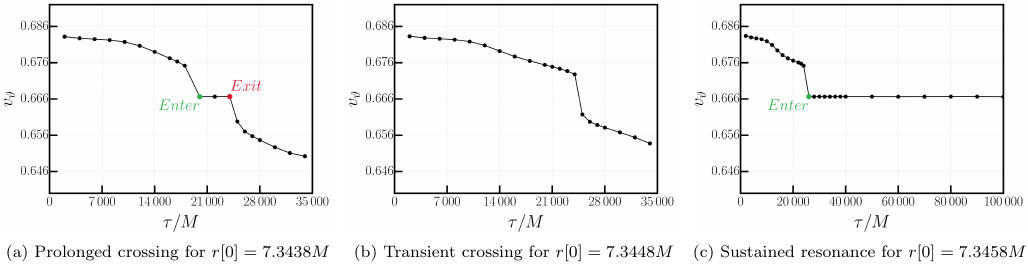}
        \caption{Rotation number along the inspiral through the 2/3 resonance, evaluated by pausing dissipation. Other initial parameters are $\mathscr{Q} = 1.0\cdot10^{-6} M^{-2}$, $E[0] = 0.98\mu$, $L_z[0] = 4.0\mu M$, $p^r[0] = 0$, $\theta[0] = \pi/2$, $\Delta \tau = 0.25M$, $\mathcal{R} = 500$, and mass ratio $q = 10^{-3}$. The fluxes were calculated directly (no interpolation) with $n = 25$ and $\mathcal{T} = 0.01M$.}
        \label{fig: General resonance crossings through 2/3}
        \end{figure*}
        
        We study inspirals that cross the 4/5 resonance in the spacetime shown in Fig.~\ref{fig: Poincare_Rotation}. Across all tested initial radii, we observe three generic behaviors:
        \begin{itemize}
            \item \emph{Prolonged resonance crossing}: the orbit enters and exits within a finite interval \cite{Lukes-Gerakopoulos2022} and the rotation number exhibits a clear plateau,
            \item \emph{Transient resonance crossing}: successive samples miss the plateau interval \cite{Flanagan2012} and only a visible discrete step in the rotation number is seen,
            \item \emph{Sustained resonance}: the orbit enters and does not leave within the simulated time \cite{vandeMeent2014}.
        \end{itemize}
        
        Entry into the resonance is accompanied by the characteristic banana-shaped structure in the Poincaré section. More decisively, the tracing of the KAM curve reverses direction (from counterclockwise to clockwise) upon entering the resonance, a typical signature of resonance crossing. Throughout the inspiral, we also observe a gradual shrinkage of the radial distance of $\mathbf{x}_s$ as the region of the bounded orbits moves towards the primary black hole.
        
        Figure~\ref{fig: General resonance crossings through 4/5} illustrates these cases for crossing $4/5$ resonance for three initial radii that differ by $\Delta r \sim 10^{-3}M$, underscoring the sensitivity to the starting point. The top row shows the rotation number versus proper time, and the bottom row shows the corresponding stroboscopic Poincaré sections. The left and right columns display prolonged resonance crossings with a 4/5 plateau, while the middle column demonstrates a sustained resonance, where the particle remains within a single island.
        
        The left panel of Figure~\ref{fig: Additional plots} shows the complete Poincaré section from the trapping case depicted in Fig.~\ref{fig: General resonance crossings through 4/5}. Because every fifth point is selected, five distinct sets can be chosen; the colors in the plot distinguish these sets.
        
        Applying the same approach to the $2/3$ resonance in the previously studied spacetime (left panel of Fig.~\ref{fig: Poincare sections}) encounters difficulties due to a non-negligible chaotic layer surrounding the resonance. In this setting, upon entering the resonance, the rotation number fluctuates irregularly, which obscures the precise entry and exit times that are important for our analysis. The right panel of Fig.~\ref{fig: Additional plots} illustrates such behavior. To solve this issue, we reduce the perturbation to $\mathscr{Q} = 1.0 \cdot 10^{-6} M^{-2}$ and study the $2/3$ resonance in a spacetime where the chaotic layer is narrower (see right panel of Fig.~\ref{fig: Poincare sections}). In that case, the discrete sampling used by both diagnostics no longer interferes with the identification of the resonance, as in the $4/5$ case discussed above.
        
        Figure~\ref{fig: General resonance crossings through 2/3} shows crossings of the 2/3 resonance. Here we present only rotation curves, because the inspiral is sufficiently fast that the stroboscopic depiction would include too few intersections to be informative. The left panel clearly shows the prolonged resonance crossing. The middle panel demonstrates a transient resonance crossing with a discrete jump in the rotation number. The right panel again shows a sustained resonance. These three qualitatively different cases arise for initial radii within the narrow interval $r \in (7.3438M,7.3458M)$, highlighting the strong sensitivity on initial conditions, i.e. on the entrance phase.
        
        An important observation is that the inspiral cannot be evolved indefinitely. All three $2/3$ cases eventually reach a regime where the motion becomes unbound due to the cumulative loss of energy and angular momentum: for the first two, around $\tau \approx 2.5\times 10^{5}M$, and for the third, bound motion persists up to $\tau = 9.6\times 10^{5}M$. The particle thus either plunges into the central black hole or escapes to infinity. Moreover, in the third case, the particle remains trapped within the resonance for the entire duration of its calculated bound motion.
        
        As noted in Fig.~\ref{fig: General resonance crossings through 2/3}, we used a non-interpolated (direct) flux evaluation. Because this resonance lies near the edge of the bound-motion domain, a dense grid would include many nonphysical nodes, reducing interpolation efficiency; a slower, non-gridded approach is, therefore, preferable in this instance.
        
        \begin{figure*}[t]\centering
        \includegraphics[width=1.0\textwidth]{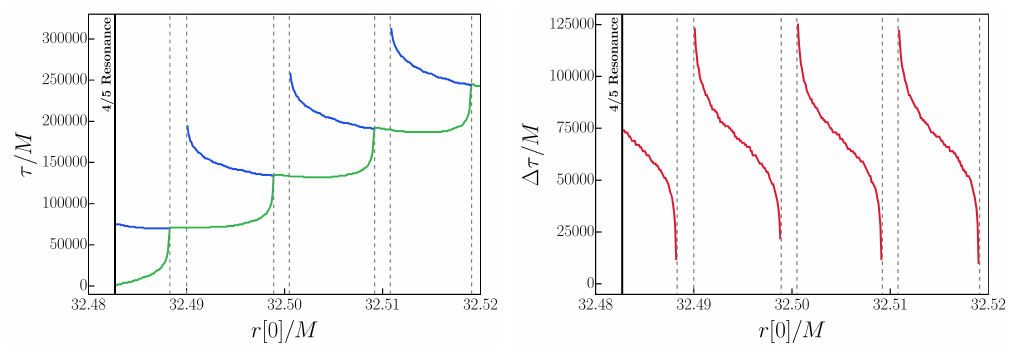}
        \caption{Left: entry (green) and exit (blue) times of particles crossing the $4/5$ resonance. Right: corresponding duration spent inside the resonance. The thick black line denotes the edge of the resonance, and gray dashed lines mark intervals with no observed exits within $\tau = 2\cdot10^{5}M$.}
        \label{fig: 4/5 entry-exit}
        \end{figure*}
        
    \subsection{Resonance crossing through 4/5 resonance}
        
        In the previous subsection, we found that there is a strong sensitivity to the initial position (phase) of the inspiral, and we observed that different types of motion are periodically repeated as we move farther from the resonance. Therefore, it is natural to study this dependence on the initial position in a systematic way.
        
        With the methodology established in the previous subsection, we perform a systematic survey of inspirals crossing the $4/5$ resonance. The initial radial coordinate is sampled in the narrow interval $r[0] \in (32.4828M; 32.5200M)$ with step $\Delta r = 10^{-4}M$, resulting in $373$ distinct adiabatic inspirals. All other parameters are fixed to ${\mathscr{Q} = 5.0 \cdot 10^{-6} M^{-2}}$, $E[0] = 0.98\mu$, $L_z[0] = 4.0\mu M$, ${p^r[0] = 0}$, $\theta[0] = \pi/2$, $\Delta\tau = 0.25M$, $\mathcal{R} = 200$, and $q = 10^{-3}$.
        
        For each inspiral, we determine the entry and exit times of the resonance. The left panel of Fig.~\ref{fig: 4/5 entry-exit} displays the detected entry (green) and exit (blue) times, while the right panel shows the duration spent inside the resonance. These times were determined with a resolution of $\delta \tau = \pm 1000 M$.
        
        The results exhibit a highly structured and recurring pattern as a function of the initial radius. Hence, there exists a continuous range of initial radii for which the qualitative behavior of the motion remains the same: precisely the three variants discussed in the previous subsection. The presence of sharp cusps in the entry and exit curves implies that there exist two trajectories that are arbitrarily close to each other: one spends effectively zero time in the resonance, while the other becomes trapped for a potentially infinite duration.
        
        These curves exhibit a step-like profile, which can be smoothed by using a finer resolution to determine the entry/exit times. However, with finer resolution (${\delta\tau = 100M}$), we encountered a small chaotic layer, which ultimately spoiled the smoother profile. Therefore, although the chaotic layer is still present, our original resolution effectively bypasses it. 
        
        Overall, this analysis demonstrates that the time spent within the $4/5$ resonance can vary widely for initial radii differing by only $\Delta r \sim 10^{-4}M$.
        
    \subsection{Resonance crossing through 2/3 resonance}
        
        Following the previous results, we construct the same plots for the $2/3$ resonance. We simulate $509$ distinct adiabatic inspirals with initial radii sampled as $r[0] \in (7.2984M;7.4000M)$ with step $\Delta r = 2\cdot10^{-4}M$, while keeping the remaining parameters identical to those used in the previous subsection: $\mathscr{Q}=1.0\cdot10^{-6}M^{-2}$, $E[0]=0.98\mu$, $L_z[0]=4.0\mu M$, $p^r[0]=0$, $\theta[0]=\pi/2$, $\Delta\tau=0.25M$, $\mathcal{R}=500$, and $q=10^{-3}$. In this scan, the fluxes are computed directly (no interpolation) with $n=25$ and $\mathcal{T}=0.01M$ to avoid possibly empty grid points. For each inspiral, the entry and exit times are determined with a resolution of $\delta\tau=\pm 500M$.
        
\begin{figure*}[t]\centering
        \includegraphics[width=1.0\textwidth]{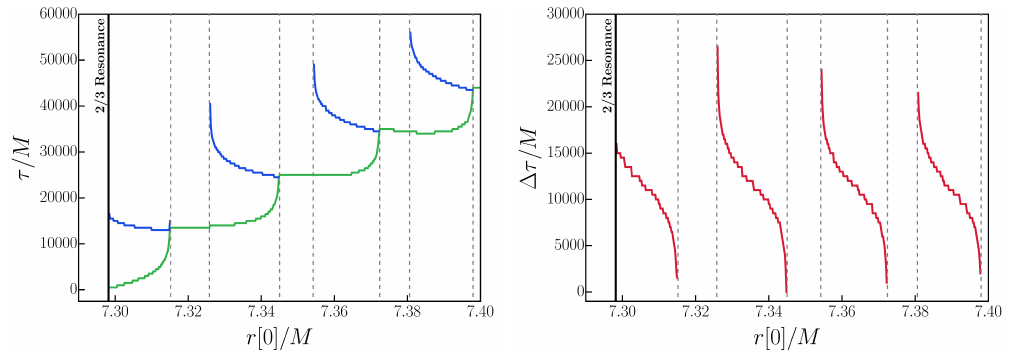}
        \caption{Left: entry (green) and exit (blue) times of particles crossing the $2/3$ resonance. Right: corresponding duration spent inside the resonance. The thick black line denotes the edge of the resonance, and gray dashed lines mark intervals with no observed exits within $\tau = 2\cdot10^{5}M$.}
        \label{fig: 2/3 entry-exit}
        \end{figure*}

        Figure~\ref{fig: 2/3 entry-exit} (left) shows the measured entry (green) and exit (blue) times; the right panel reports the corresponding time spent inside the resonance. Qualitatively, the structure mirrors the $4/5$ case: intervals of prolonged resonance crossing  alternate with trapping windows of sustained resonances, separated by sharp cusps in which neighboring initial data yield drastically different times spent in the resonance. As before, step-like features reflect the finite timing resolution used for entry/exit identification.
        
        A final remark concerns the magnitude of the quadrupole parameter used across our simulations. The values were chosen to be large to make resonance effects clearly visible. For more realistic rings (${r_r \sim 100M}$, ${\mathscr{M}_r \sim 0.1M}$), the effective $\mathscr{Q}$ would be smaller, narrowing the resonance and shortening the resonance crossing times. The present results should therefore be interpreted as qualitative templates of the dynamical behavior rather than precise predictions for specific astrophysical configurations.
        
        \subsection{Dependence on the mass ratio}
        
\begin{figure*}[t]\centering
        \includegraphics[width=1.0\textwidth]{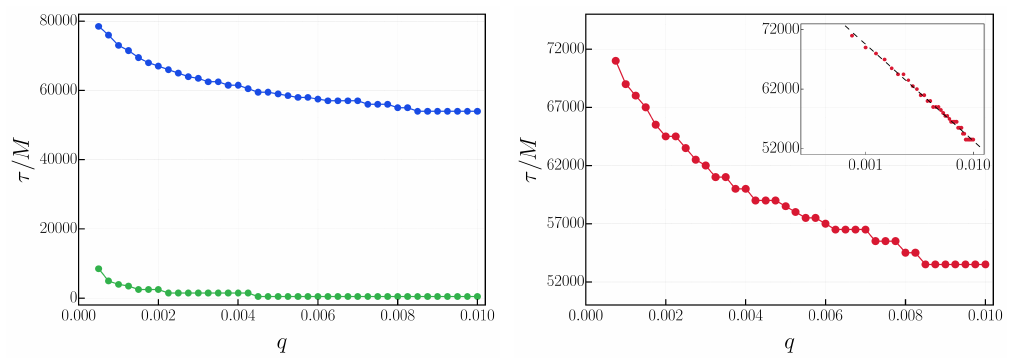}
        \caption{Left: entry (green) and exit (blue) times of particles crossing the $4/5$ resonance. Right: corresponding duration spent inside the resonance. The inset plot of the bottom panel shows the duration spent in resonance on a logarithmic scale with a linear fit.}
        \label{fig: mass ratio entry-exit}
        \end{figure*}

        In all previous adiabatic inspirals, we have used an identical mass ratio $q = 10^{-3}$. To assess how these results scale with the system parameters, we vary the mass ratio $q$ while keeping the initial radius fixed at $r[0] = 32.4840M$ (other parameters are identical to those in Fig.~\ref{fig: General resonance crossings through 4/5}). For each inspiral, we again determine resonance entry and exit by the same bisection-based procedure used above, with a resolution $\delta \tau = \pm 1000M$. Results are shown in Fig.~\ref{fig: mass ratio entry-exit}.

        The time spent inside the resonance within the $4/5$ resonance decreases approximately exponentially with increasing $q$, evidenced by the linear trend on a logarithmic scale. Therefore, $q$ acts as a global timescale parameter for the inspiral, justifying our baseline choice $q = 10^{-3}$ as a compromise between observable time inside the resonance and computational cost. Moreover, a more important conclusion from this test is that as $q \rightarrow 1$ (comparable-mass systems), the time inside the resonance becomes negligible, rendering resonance effects dynamically irrelevant for such binaries, which agrees with existing conjectures \cite{Hughes2000}. However, when $q$ approaches the values relevant for EMRI systems, the treatment of resonances becomes a subtle point.
        
        One more test was conducted, where the previous setting was retained but the initial radius was moved farther from the resonance to $r[0] = 32.4910M$. This test again showed a decreasing dependency on mass ratio, however, it also showed that the whole cotangent-like profile from Fig.~\ref{fig: General resonance crossings through 4/5} shifts to the left. The mass ratio $q$ not only rescales the clock, but also modifies flux amplitudes locally and therefore slightly shifts the inspiral in phase space, which in turn affects the location and duration of the crossing.
        
        In summary, the mass ratio primarily governs how long an inspiral can interact with a resonance, and secondarily, perturbs the location where interaction occurs in phase space. These results reinforce the expectation that resonant phenomena are most consequential for EMRIs and progressively less so toward comparable-mass binaries with $q \sim 1$.
        
\section{Conclusion}\label{sec: Section6}
        We presented a systematic study of gravitational-wave fluxes and resonance crossings for EMRI orbits in a perturbed Schwarzschild spacetime with a ring perturbation truncated at its quadrupole moment. First, we benchmarked three computational schemes for fluxes: quadrupole (QP), post-Newtonian (PN), and Teukolsky-based (TK), across unperturbed and perturbed settings. In the unperturbed regime, the methods agree, whereas in the perturbed near-integrable regime PN loses validity and QP remains in qualitative and quantitative agreement with TK. A decomposition of the TK fluxes established $m=\pm2$ as the dominant azimuthal modes, with subdominant modes shared primarily by $m=\pm1$ and $m=\pm3$. Although the TK method incorporates relativistic corrections beyond leading order, the computational cost is burdensome. Nevertheless, these results provide a validated foundation for using the QP fluxes in subsequent resonance-crossing analyses.
        
        To enable large-scale inspiral scans, we introduced two numerical accelerants: a robust interpolation of precomputed fluxes and a repetitiveness parameter that updates fluxes sparsely along the trajectory. Together they deliver an overall speedup of about $500\times$ while keeping trajectory errors within adiabatic tolerances. Building on this infrastructure, we generated dense families of inspirals crossing the $2/3$ and $4/5$ resonances and mapped entry/exit times and time spent inside the resonance. The scans reveal the three known generic behaviors: prolonged resonances \cite{Lukes-Gerakopoulos2022}, transient resonances \cite{Flanagan2012} with discrete steps in the rotation number, and sustained resonances \cite{vandeMeent2014}. These results show a strong sensitivity to the initial radius.
        
        Across both the cases of resonances we examined, we find structured ranges in initial radius where the qualitative behavior is stable, separated by sharp cusps and jumps where adjacent trajectories experience widely different times inside the resonance. Increasing $q$ primarily shortens the timescale of the resonance crossing, rendering resonant effects progressively less significant as one moves away from the EMRI limit toward comparable-mass binaries. These conclusions align with earlier indications from Ref.~\cite{Hughes2000}.
        
        In summary, our work demonstrates the agreement between the QP and TK in the perturbed near-integrable regime, presents a scalable computational pipeline for adiabatic inspirals, and provides a high-resolution phenomenology of resonance crossings, including trapping in sustained resonances, and with $q$-scaling of resonance crossing times.

\section*{Acknowledgments}\label{sec: Section7}
        MS, GLG, and OZ have been supported by the fellowship Lumina Quaeruntur No. LQ100032102 of the Czech Academy of Sciences. OZ has also been supported by the PPLZ fellowship of the Czech Academy of Sciences.

    \bibliographystyle{unsrt}
    \bibliography{Bibliography}

\appendix
    \section{Tolerance test}\label{sec: Tolerance test}
        
        In the numerical study of fluxes, it is necessary to define the appropriate endpoint of a trajectory’s evolution since fluxes are averaged along the entire trajectory. We employ a \textit{tolerance test} to determine when the motion has effectively completed full "period"\footnote{Here, "period" denotes a near return to the initial radial position, acknowledging the motion is not exactly periodic in general.}.  
        
        This method uses two parameters:  
        \begin{itemize}
            \item the number $n$ of revolutions, counted when the trajectory reaches radial extrema of the same type (minimum or maximum) as the initial condition\footnote{Since each trajectory begins with $p^r[0] = 0$, the motion initiates in radial extrema.},  
            \item the distance $\mathcal{T}$ between the initial radial position and subsequent extrema.  
        \end{itemize}
        
\begin{figure}[t]\centering
        \includegraphics[width=0.45\textwidth]{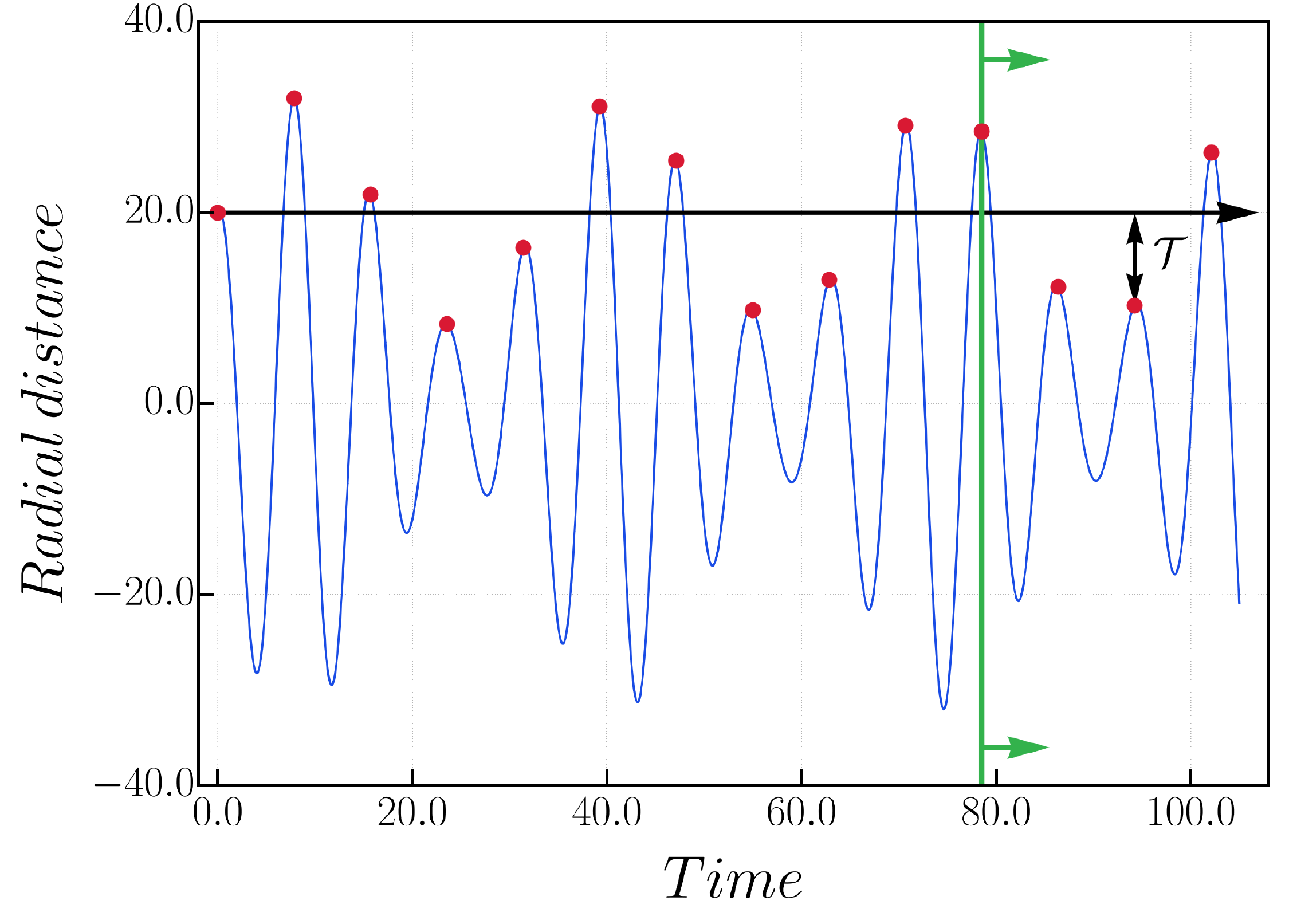}
        \caption{
        Illustration of the tolerance test. The blue curve shows the radial distance, red dots mark radial maxima, the green line indicates the point where the minimal number of revolutions ($n=10$ in this example) is reached, and $\mathcal{T}$ is the distance between the initial radius and subsequent extrema.
        }
        \label{fig: Tolerance_test_illustration}
        \end{figure}

        Once the trajectory has completed at least $n$ prescribed revolutions, the test checks $\mathcal{T}$ at each relevant extremum. The evolution is terminated when $\mathcal{T}$ falls below the chosen threshold (see Fig.~\ref{fig: Tolerance_test_illustration}).
        
        \begin{figure*}[t]\centering
        \includegraphics[width=1.0\textwidth]{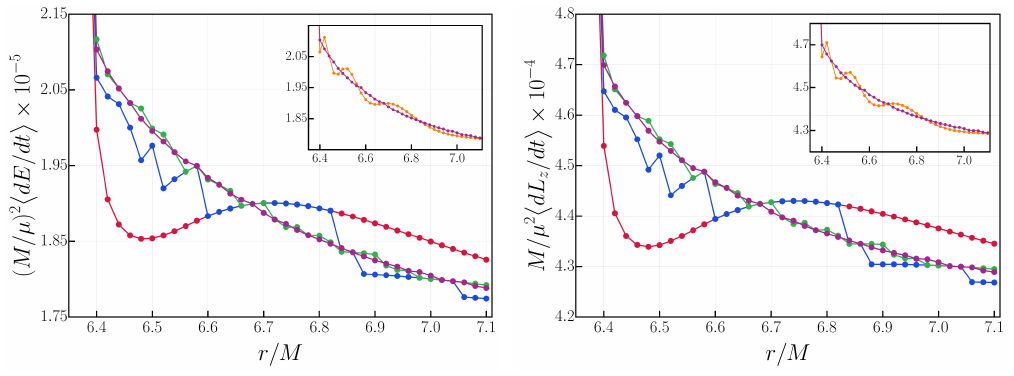}
        \caption{
        QP fluxes near the edge of $2/3$ resonance. Left panel shows the energy fluxes and right panel the angular momentum fluxes. All curves use identical parameters: $\mathscr{Q} = 2.5\cdot10^{-6} M^{-2}$, $E=0.98\mu$, $L_z=4.00\mu M$, $\theta[0]=\pi/2$, $p^r[0]=0$, $r[0]\in(6.30M,7.10M)$ with step $\Delta r=0.02M$ and time step $\Delta \tau=0.05M$. The red curve uses fixed $n=10$ with no tolerance parameter ($\mathcal{T}\to\infty$); the blue, green, and purple curves use $n=10$ with $\mathcal{T}=0.1M$, $0.01M$, and $0.001M$, respectively. In the inset, the orange curve corresponds to fixed $n=50$.}
        \label{fig:Tolerance_test_resonance}
        \end{figure*}
        
        \begin{figure*}[t]\centering
        \includegraphics[width=1.0\textwidth]{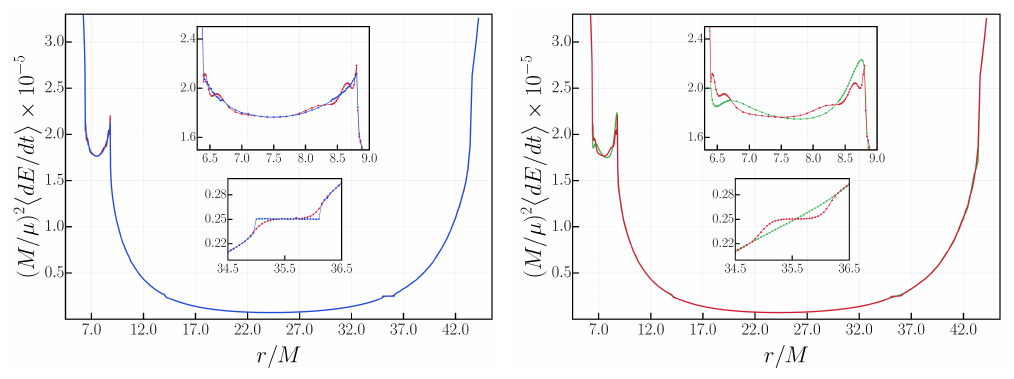}
        \caption{
        Averaged QP energy fluxes across the full radial domain. 
        Parameters: $\mathscr{Q}=2.5\cdot10^{-6}M^{-2}$, $E=0.98 \mu$, $L_z=4.00 \mu M$, $\theta[0]=\pi/2$, $p^r[0]=0$, $r[0]\in(6.2M,44.2M)$ with $\Delta r=0.2M$ (refined near resonances). 
        Left: red curve uses fixed $n=33$, blue curve has $n=10$ with $\mathcal{T}=0.005M$ ($n_\text{av}=33$). 
        Right: red curve stays same, green curve is for fixed $n=10$. 
        Insets zoom on the $2/3$ (top) and $4/5$ (bottom) resonances.}
        \label{fig:Tolerance_test_whole}
        \end{figure*}        
        
        The parameter $n$ ensures a minimum integration time for meaningful averaging, which depends on the specific orbital configuration. The tolerance $\mathcal{T}$ determines when the trajectory sufficiently returns to its initial radial distance, providing an effective definition of a "period". Since the perturbed motion is non-periodic (in general) and does not return exactly to its initial phase space position, the tolerance $\mathcal{T}$ effectively defines how close is "close enough" to be considered a full return.
        
        We demonstrate this method on fluxes near the $2/3$ resonance. In Fig.~\ref{fig:Tolerance_test_resonance}, we show GW fluxes computed with the QP formalism for several values of $n$ and $\mathcal{T}$. The purple curve exhibits the anticipated smooth behavior of fluxes, while other curves show distortions due to insufficient averaging.  
        
        The average number of revolutions highlights the effect: for the red curve $n_\text{av}=10$, for the blue $n_\text{av}=13$, and for the green $n_\text{av}=23$. Therefore, a minor increase between red and blue in computational time significantly affects the result. Moreover with green setting, we get almost the correct profile.
        
        In the inset, the purple curve with $n_\text{av}=53$ achieving better accuracy than the orange one with $n_\text{av}=50$, showing the effectiveness of tolerance test with nearly the same computational cost. For the purple curve, the number of revolutions varies significantly between trajectories: thirteen required fewer than 30 revolutions (one with only $n=10$), while five needed more than 100. Thus, the total computational cost is not reduced but redistributed, with some trajectories requiring above-average resources and others below.
        
        Note that both energy and angular momentum fluxes show comparable sensitivity to the tolerance test.
        
        Importantly, the number of revolutions is independent of a trajectory’s position relative to the resonance: some low-revolution cases lie near the resonance edge, while some high-revolution ones occur outside or within it. The tolerance test therefore serves only to define the termination of the evolution and no additional dynamical information should be inferred from it.
        
        Figure~\ref{fig:Tolerance_test_whole}, panel (a), presents a full scan over the initial radial range. With 280 trajectories and an average of $n\approx 33$ revolutions, both curves have identical computational cost, allowing a direct comparison. Far from resonances, the tolerance test is largely irrelevant; near resonances, however, it is essential. The red curve fails to reproduce the expected symmetry of the flux profile about the resonant island’s fixed point~\cite{Strateny2023b,Strateny2023a}, while the blue curve satisfies it.
        
        Furthermore, for the 4/5 resonance, the evolution without the tolerance test barely reflects the resonance, while evolution with it clearly reveals the characteristic symmetric plateaus (see Ref.~\cite{Strateny2023b,Strateny2023a}).  
        
        The effect of $n$ is shown in Fig.~\ref{fig:Tolerance_test_whole}, panel (b): without the tolerance test, the flux profile converges only slowly with increasing revolutions, so $n$ must be sufficiently large to obtain meaningful results. For QP fluxes this is not problematic, as the computation is efficient even on an average laptop. In contrast, for TK fluxes, a single revolution on a reasonable grid costs about 5 CPU core-hours. For example, during the $2/3$ resonance simulations the Teukode ran on 25 cluster cores with a $801\times61$ grid, producing only $4$–$5$ revolutions per hour. Hence, minimizing the number of revolutions is essential for Teukode simulations.
        
\begin{figure}[h]\centering
        \includegraphics[width=0.45\textwidth]{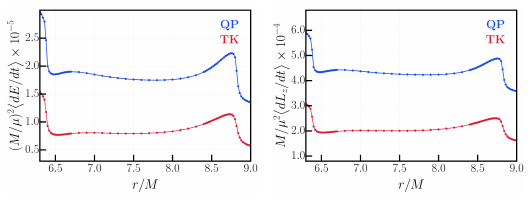}
        \caption{The QP fluxes (blue) and TK fluxes for $m=2$ using a grid size of $801 \times 61$ (red). Left panel shows energy fluxes and right panel shows the angular momentum fluxes. Parameters: $\mathscr{Q} = 2.5\cdot10^{-6} M^{-2}$, $E = 0.98 \mu$, $L_z=4.00 \mu M$, $\theta[0] = \pi/2$, $p^r[0] = 0$, $r[0] \in (6.2M, 9.0M)$ with step size $\Delta r = 0.1 M$, with refinement around the resonance edge to $\Delta r = 0.02M$. Both curves are without the tolerance test, i.e., with fixed $n = 10$.}
        \label{fig:Tolerance_test_TK}
        \end{figure}

        Finally, note that the tolerance test has so far been applied only to QP fluxes. In early Teukode runs on resonant cases (without the test), we observed same effect as for QP fluxes, which is shown in Fig.~\ref{fig:Tolerance_test_TK}.

    \section{Equatorial motion in the Schwarzschild spacetime} \label{sec:EquatFlux}

        \begin{figure*}[t]\centering
        \includegraphics[width=1.0\textwidth]{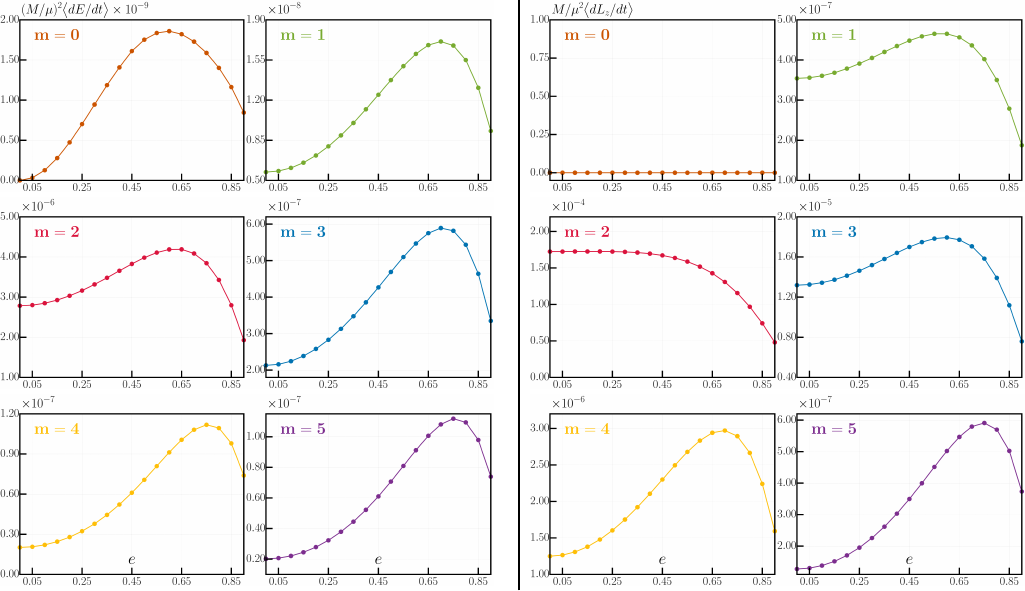}
        \caption{Individual TK $m$ mode contributions to the energy (left two columns) and angular momentum (right two columns) fluxes for the equatorial motion in the Schwarzschild spacetime as a function of eccentricity.}
        \label{fig: Equatorial motion - m-modes}
        \end{figure*}

        \begin{figure*}[t]\centering
        \includegraphics[width=1.0\textwidth]{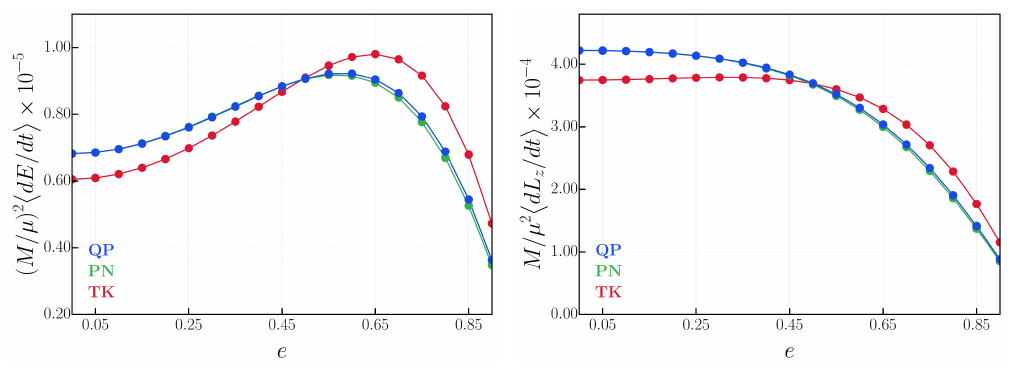}
        \caption{Comparison of energy (left) and angular momentum (right) fluxes computed using the three different methods for the equatorial motion in the Schwarzschild spacetime.}
        \label{fig: Equatorial motion - full plot}
        \end{figure*}
        
        This appendix summarizes the baseline comparison of the three computational methods for equatorial motion in Schwarzschild spacetime. The results collected here are referenced in the main text and are intended to illustrate the key properties and differences among the approaches.
        
        All orbital parameters are fixed: $\mathscr{Q} = 0$, $L_z = 4.40\mu M$, $p^r[0] = 0$, $\theta[0] = \pi/2$, $\dot{\theta}[0] = 0$. The eccentricity is varied by adjusting $r[0]$, while the inclination remains zero ($\iota \equiv 0$). The orbital energy $E$ is determined from four-momentum conservation.
        
        The Teukode simulations were performed on a grid of $1001 \times 76$ points. Since the motion is strictly periodic, the tolerance test was unnecessary; fluxes were averaged over $n = 10$ revolutions.
        
        Figure~\ref{fig: Equatorial motion - m-modes} shows the decomposition of energy and angular momentum fluxes into azimuthal modes according to Eq.~\eqref{eq: Decomposition into m-modes}. The $m = \pm 2$ and $m = \pm 3$ modes dominate, together contributing more than $95\%$ of the total flux across all eccentricities.
        
        Summing over the individual modes allows for a direct comparison with the QP and PN results. As shown in Fig.~\ref{fig: Equatorial motion - full plot}, the QP and PN fluxes are in excellent agreement. The TK method reproduces the same qualitative dependence on eccentricity, although minor systematic differences in magnitude remain. This pattern is consistent with the tests presented in the main text.
        
        For completeness, we also performed simulations of inclined motion in Schwarzschild spacetime, where the orbital plane is tilted and the trajectory evolves in the $\theta$ direction as well. The results remain qualitatively similar to the equatorial case: the QP and PN methods agree closely, while the TK fluxes follow the same overall trend with slightly different magnitude.
        
        The more intriguing behavior arises when perturbations are introduced to the Schwarzschild spacetime via an external quadrupole, as we discussed in the main text.
        
    \section{Parametrization of the trajectory}\label{sec: Parametrization of the trajectory}

        \begin{figure}[h]\centering
        \includegraphics[width=0.45\textwidth]{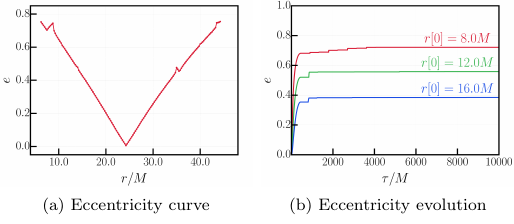}
        \caption{
        Left: eccentricity curve for identical parameters, namely $\mathscr{Q} = 2.5\cdot10^{-6}M^{-2}$, $E = 0.98 \,\mu$, $L_z = 4.00 \,\mu M$, $\theta[0] = \pi/2$, $p^r[0] = 0$, and $r[0] \in (6.25M,\,44.25M)$ with step size $\Delta r = 0.25M$. Right: time evolution of three sample initial radii from the left panel, illustrating convergence of $e(\tau)$.
        }
        \label{fig:Eccentricity curves}
        \end{figure}
        
        As mentioned in the main text, we parametrize the trajectory by three parameters $(E, L_z, e)$. The first two are straightforward, as they remain conserved along geodesics, whereas the third must be chosen. In the unperturbed system one may use $L^2$, or equivalently the Carter constant, to complete the set. However, once the perturbation is included, the integral of motion $L^2$ is no longer available. Therefore, after several tests, we adopt the eccentricity as the third parameter. This parameter is easy to determine for each geodesics, since it only requires tracking the minimal and maximal radial distance. In practice, we estimate $e$ from the turning points $(r_{\min}, r_{\max})$ over a short evolution and then use it for interpolation.
        
        This choice is required to satisfy two practical conditions: (i) it must be computable sufficiently fast so that flux interpolation is cheaper than direct flux evaluation, and (ii) it must map each eccentricity value to the correct trajectory (and hence to the correct fluxes).
        
        The first condition is illustrated in the right panel of Fig.~\ref{fig:Eccentricity curves}. There we observe that the eccentricity converges to its actual value after $\tau \approx 5000M$. Consequently, evaluating $e$ on the current geodesic at each step and then interpolating the fluxes still reduces the overall computational cost compared to direct flux computation.
        
        The second condition is illustrated in the left panel of Fig.~\ref{fig:Eccentricity curves}, which shows a full scan of eccentricities for a fixed pair $(E,L_z)$. The eccentricity curve resembles a V-shaped curve centered around the middle of the main island of stability; similar V-shaped structures appear around resonances. Knowing the relative position with respect to the center (i.e., whether the trajectory lies on the left or right branch) enables an unambiguous mapping between the initial radius and the eccentricity, and vice versa. Throughout the inspiral we assume this relative position remains unchanged. The uniqueness of this mapping breaks down in the immediate vicinity of a resonance, where two distinct initial radii can share the same eccentricity. However, as verified by our tests of GW fluxes near resonances (see Sec.~\ref{sec: Resonant motion}), these two radii yield identical fluxes. Thus, although $e$ does not uniquely determine the initial radius within a resonance, the resulting ambiguity is irrelevant for flux value.

    \section{Interpolation}\label{sec: Interpolation}
        
        To accelerate the adiabatic inspiral computation, the GW fluxes were precomputed on a discrete three-dimensional grid of parameters $(E, L_z, e)$ and interpolated during evolution. This approach replaces repeated direct flux evaluations with a numerically efficient interpolation procedure.
        
        The interpolation task is complicated by two factors. First, the grid is irregular along the eccentricity dimension, since $e$ is derived from the initial radius $r[0]$, and its values are not uniformly spaced for different combinations of $(E, L_z)$. This irregularity is unavoidable in our setting and complicates direct gridding in $e$. Second, certain grid points correspond to non-physical or plunging/escaping trajectories, for which fluxes are undefined and must be excluded. Consequently, the interpolation must handle irregular and partially incomplete data.
        
        We tested two methods for the interpolation: \textit{barycentric interpolation} and \textit{cubic-spline interpolation}. The barycentric method is a polynomial-based approach that is fast and simple, but it can suffer from numerical instabilities for dense and equally spaced grids. The cubic-spline method, in contrast, provides a smooth interpolation and remains stable even for irregular data, but with greater numerical cost. In our implementation, spline interpolation was performed using the GNU Scientific Library (GSL, \cite{Galassi2019}), and both approaches were benchmarked against \textit{Mathematica}'s internal interpolation to ensure consistency.
        
        The final interpolation scheme follows a simple dimensional reduction strategy. For each fixed pair $(E, L_z)$, the fluxes are first interpolated along $e$, then along $L_z$, and finally along $E$, both for $dE/dt$ and $dL_z/dt$. This approach allows accurate flux estimation at any point within the grid while keeping the computational cost low during the inspiral evolution.
        \newline
        \subsection*{Interpolation tests}
        To verify the reliability of the interpolation, we performed several controlled tests on both artificial and physical datasets.

        \begin{figure}[t]\centering
        \includegraphics[width=0.45\textwidth]{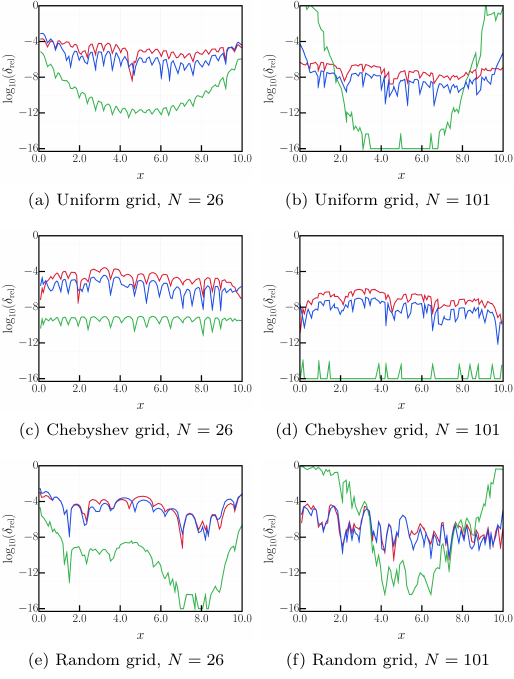}
        \caption{
        Relative errors of interpolation methods for a $1D$ test function on uniform, Chebyshev and random grids. The red curve corresponds to Mathematica, green to barycentric, and blue to cubic-spline interpolation, $N$ denotes number of nodes.
        }
        \label{fig:1D_interp_test}
        \end{figure}        
        
        \paragraph{1D test.}
        We first compared all three interpolation methods (barycentric, cubic-spline, and Mathematica) using a smooth one-dimensional test function designed to mimic the behavior of flux profiles near resonances. For brief results, see Fig.~\ref{fig:1D_interp_test}.  
        On an equally spaced grid, barycentric interpolation showed large oscillations near the interval edges for denser grids (the Runge phenomenon). When Chebyshev nodes were used, the Runge effect disappeared and barycentric interpolation became significantly more accurate, reaching near machine precision for dense grids, outperforming the other two methods. When the data were sampled randomly across the interval, the barycentric interpolation becomes unstable, with errors exceeds unity in some regions. This breakdown is expected, as barycentric interpolation assumes structured grids and lacks stability for arbitrarily spaced nodes. By contrast, cubic splines and Mathematica remained consistent and reliable across all tested densities.

        \begin{figure}[!t]\centering
        \includegraphics[width=0.45\textwidth]{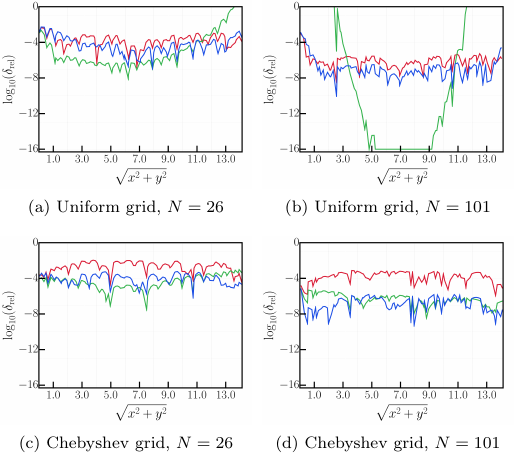}
        \caption{
        Relative errors of interpolation methods for a $2D$ test function on uniform and Chebyshev grids. The red curve corresponds to Mathematica, green to barycentric, and blue to cubic-spline interpolation, $N$ denotes number of nodes.
        }
        \label{fig:2D_interp_test}
        \end{figure}
        
        \paragraph{2D test.}
        
        Next, we extended the comparison to a two-dimensional test function, representing the $(E, L_z)$ plane for a fixed eccentricity. Here, we no longer assume randomly spaced nodes, since we can always control sampling in the $(E, L_z)$ plane. For results, see Fig.~\ref{fig:2D_interp_test}. On dense, $2D$ equally spaced grids is Runge’s phenomenon intensified and barycentric interpolation became unstable, producing large errors. On other hand, cubic splines and Mathematica interpolation remained accurate. When Chebyshev spacing was used, the performance of the barycentric method improved and oscillations were removed.  However, even under optimal Chebyshev spacing, barycentric interpolation did not outperform cubic splines in 2D; both were accurate, with splines remaining the more robust choice for higher dimensions.
        
        This $2D$ test demonstrates that when interpolation methods are generalized
        to higher dimensions, cubic spline interpolation is the most robust and accurate among the three methods evaluated. Even with the Chebyshev grid designed explicitly for barycentric interpolation, the cubic spline interpolation is not outperformed. This validates using cubic spline interpolation in all three dimensions without concern for Runge’s phenomenon, even at high grid densities.
        
        \paragraph{Flux-grid benchmark.}
        
\begin{figure}[t]\centering
        \includegraphics[width=0.45\textwidth]{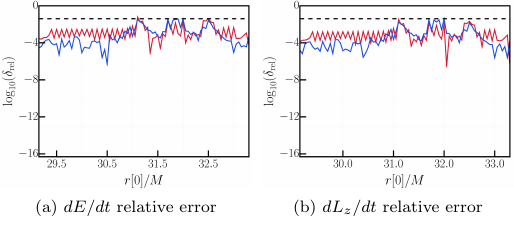}
        \caption{
        The comparison of the relative error of two interpolation methods with respect to the real values of the fluxes. Red curves show Mathematica results, blue curves correspond to the cubic-spline interpolation. The dashed line marks a $4\%$ error level.
        }
        \label{fig:flux_interp_test}
        \end{figure}

        Finally, we test the barycentric method on a real $3D$ grid of gravitational-wave fluxes computed around the $4/5$ resonance. We compared interpolated fluxes at off-grid points with directly computed reference values. For results, see Fig.~\ref{fig:flux_interp_test}. The cubic-spline interpolation achieved average relative errors of $4.0\times10^{-3}$ for $dE/dt$ and $2.8\times10^{-3}$ for $dL_z/dt$, with the largest deviations below $4\%$.

        \section{Repetitiveness parameter}\label{sec: Repetitiveness parameter}
        
        As shown in Appendix~\ref{sec: Interpolation}, the flux interpolation provides a stable and accurate method for evaluating the adiabatic inspiral. However, the high computational cost still limits the achievable evolution time, especially for multiple resonance crossings. To further accelerate the computation, we introduce the repetitiveness parameter~$\mathcal{R}$.
        
        The adiabatic inspiral evolves through a sequence of quasi-periodic geodesics, whose parameters $(E, L_z, e)$ vary only slightly between steps. Consequently, the associated GW fluxes $\dot{E}$ and $\dot{L}_z$ also change smoothly, allowing us to keep them fixed over several integration steps without losing accuracy. The repetitiveness parameter $\mathcal{R}$ defines how often the fluxes are recalculated during the evolution: $\mathcal{R} = 1$ corresponds to recomputing at every step, while $\mathcal{R} = 100$ means updating every 100th step. For a step size $\Delta \tau$, this implies recalculation whenever $\tau$ is a multiple of $\mathcal{R} \, \Delta \tau$.
        
\begin{figure}[!t]\centering
        \includegraphics[width=0.45\textwidth]{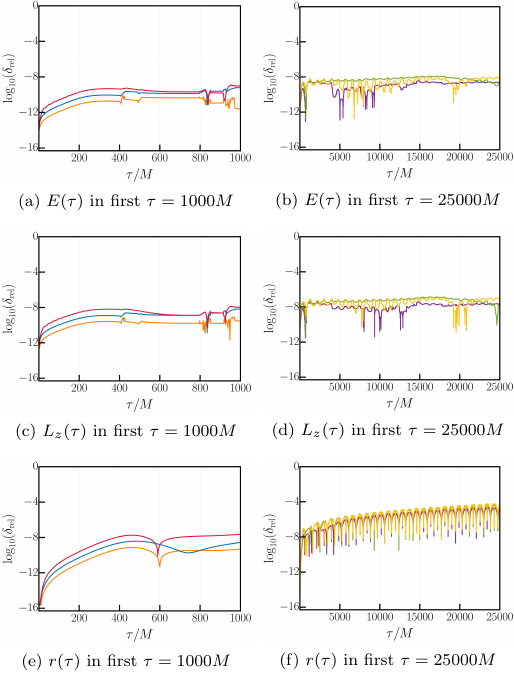}
        \caption{
        The comparison of the evolution of three orbital parameters. Initial setup: $\mathscr{Q} = 5.0\cdot10^{-6} M^{-2}$, $E[0] = 0.98\mu$, $L_z[0] = 4.0\mu M$, $r[0] = 32.4840M$, $p^r[0] = 0$, $\theta[0] = \pi/2$, $\Delta \tau = 0.25M$, $n = 10$, $\mathcal{T} = 0.01M$ and mass ratio $q = 10^{-3}$. Left: relative errors with respect to the $\mathcal{R}=1$ (orange: $\mathcal{R}=5$, blue: $\mathcal{R}=20$, red: $\mathcal{R}=100$). Right: relative errors with respect to the $\mathcal{R}=100$ (purple: $\mathcal{R}=200$, green: $\mathcal{R}=500$, yellow: $\mathcal{R}=1000$).
        }
        \label{fig:rep_test}
        \end{figure}

        To quantify the introduced error, we evolved identical trajectories for various values of~$\mathcal{R}$ (see Fig.~\ref{fig:rep_test}). Even for $\mathcal{R} = 1000$, the relative errors in energy and angular momentum remained below~$10^{-7}$, and the radial coordinate error below~$10^{-4}$, which is comparable to the integration error of the employed 4th-order Runge–Kutta scheme. Hence, the impact of~$\mathcal{R}$ on the accuracy is negligible within the adiabatic approximation.
        
        The computational gain, however, is substantial. For a trajectory evolved over $\tau = 1000M$, the runtime decreases from approximately 24 minutes for $\mathcal{R} = 1$ to 3.8 minutes for $\mathcal{R} = 5$, and only 12~seconds for $\mathcal{R} = 100$. Thus, the total runtime scales inversely with~$\mathcal{R}$, making it an effective control parameter for balancing precision and efficiency. Moreover, with sufficiently large~$\mathcal{R}$, one may even evolve the inspiral using directly computed fluxes, avoiding interpolation altogether.

\end{document}